\documentclass{article}
\pdfoutput=1
\usepackage{amsmath}
\usepackage{amssymb}
\usepackage{graphicx}
\usepackage{color}
\usepackage{soul}

\pdfpageheight\paperheight
\pdfpagewidth\paperwidth

\usepackage{jcappub}

\renewcommand{\[}{\begin{equation}}
\renewcommand{\]}{\end{equation}} 
\usepackage{array}

\begin{document}
\title{The trouble with $H_0$}
\author[a,b]{Jos\' e Luis Bernal}
\author[a,c,d,e,f]{Licia Verde}
\author[g,h]{Adam G. Riess} 

\affiliation[a]{ICC, University of Barcelona, IEEC-UB, Mart\' i  i Franqu\` es, 1, E08028
Barcelona, Spain}
\affiliation[b]{Dept. de  F\' isica Qu\` antica i Astrof\' isica, Universitat de Barcelona, Mart\' i  i Franqu\` es 1, E08028 Barcelona,
Spain}
\affiliation[c]{ICREA, Pg. Llu'\' is Companys 23, 08010 Barcelona, Spain}

\affiliation[d]{Radcliffe Institute for Advanced Study, Harvard University, MA 02138, USA}
\affiliation[f]{Institute for Theory and Computation, Harvard-Smithsonian Center for Astrophysics, 60 Garden Street, Cambridge, MA 02138, USA}

\affiliation[e]{Institute of Theoretical Astrophysics, University of Oslo, 0315, Oslo, Norway}
\affiliation[g]{Department of Physics and Astronomy, Johns Hopkins University, Baltimore, MD 21218}
\affiliation[h]{Space Telescope Science Institute, 3700 San Martin Drive, Baltimore, MD 21218}

\emailAdd{joseluis.bernal@icc.ub.edu}
\emailAdd{liciaverde@icc.ub.edu}
\emailAdd{ariess@stsci.edu}

\abstract{
We perform a comprehensive cosmological study of the $H_0$ tension between the direct local  measurement  and  the model-dependent value  inferred from the Cosmic Microwave Background. With the recent measurement of $H_0$ this tension has raised to more than $3 \,\sigma$. 
We consider changes in the early time physics without modifying the late time cosmology. 
We also reconstruct the late time expansion history in a model independent way with minimal assumptions using  distance measurements from Baryon Acoustic Oscillations and Type Ia Supernovae, finding that  at $z<0.6$ the recovered  shape of the expansion history is less than 5\% different than that of a standard $\Lambda$CDM model.  These  probes  also provide a model insensitive constraint on the  low-redshift standard ruler, measuring directly the combination  $r_{\rm s}h$  where $H_0=h \times 100$ Mpc$^{-1}$km/s   and $r_{\rm s}$ is the sound horizon at radiation drag (the standard ruler), traditionally constrained by CMB observations. Thus $r_{\rm s}$ and $H_0$ provide absolute scales for distance measurements (anchors) at opposite ends of the observable Universe.
We calibrate the cosmic distance ladder and obtain a model-independent determination of the standard ruler for acoustic scale, $r_{\rm s}$. The tension in $H_0$ reflects a mismatch between our determination of $r_{\rm s}$ and its standard, CMB-inferred value.
Without  including high-$\ell$ Planck CMB  polarization data (i.e., only considering the ``recommended baseline"  low-$\ell$ polarisation and temperature and the high $\ell$ temperature data), a modification of the early-time physics to include a component of dark radiation with an effective number of species around 0.4 would reconcile the CMB-inferred constraints, and the  local $H_0$  and  standard ruler determinations. The inclusion of the ``preliminary" high-$\ell$ Planck  CMB polarisation data disfavours this solution.}
\maketitle

\hypersetup{pageanchor=true}

\section{Introduction}\label{sec:Introduction}

In the last few years,  the determination of cosmological parameters  has reached astonishing and unprecedented precision. Within the  standard $\Lambda$ - Cold Dark Matter ($\Lambda$CDM) cosmological model some parameters are constrained at or below the percent level. This model assumes a spatially flat cosmology and matter content dominated by cold dark matter but with total matter energy density dominated by a cosmological constant, which drives a late time accelerated expansion.
Such precision has been driven by a major observational effort. This is especially true in the case of Cosmic Microwave Background (CMB) experiments, where WMAP \citep{HinshawWMAP_13,BennettWMAP_13} and Planck \citep{Planckparameterspaper} have played a key role, but also in the measurements of Baryon Acoustic Oscillations (BAO) \citep{Anderson14_bao,Cuesta16_bao}, where the  evolution of the cosmic distance scale is  now measured with a $\sim  1\%$ uncertainty.

The Planck Collaboration 2015 \citep{Planckparameterspaper} presents the strongest constraints so far in key parameters, such as  geometry,  the predicted Hubble constant, $H_0$, and the sound horizon at radiation drag epoch, $r_{\rm s}$. These last two quantities provide an absolute scale for distance measurements at opposite ends of the observable Universe (see e.g., \cite{Cuesta:2014asa, Aubourg_2015}), which makes them essential to build the distance ladder and model  the expansion history of the Universe. However, they are {\it indirect} measurements and as such  they are model-dependent. Whereas the $H_0$ constraint  assumes an expansion history model (which heavily relies  on late time physics assumptions such as the details of late-time cosmic acceleration, or equivalently, the properties of dark energy), $r_{\rm s}$ is a derived parameter which relies on early time physics (such as the density and equation of state parameters of the different species in the early universe). 

This is why having model-independent, direct measurements of these same quantities is of utmost importance. In the absence of significant systematic errors, if the standard cosmological model is the correct model,  indirect (model-dependent) and direct (model-independent) constraints on these parameters should agree. If they are significantly inconsistent, this will provide evidence of physics beyond the standard model (or unaccounted  systematic errors).

Direct measurements of $H_0$ rely on 
the ability to measure absolute distances to $>100$ Mpc, usually through the use of coincident geometric and relative distance indicators. 
$H_0$ can be interpreted as  the normalization of the Hubble parameter, $H(z)$, which describes the expansion rate of the Universe as function of redshift. Previous constraints on $H_0$ (i.e. \citep{Riess:2011yx}) are consistent with the final  results from the  WMAP mission, but  are in $2$-$2.5 \sigma$ tensions with Planck when $\Lambda$CDM model is assumed \citep{MarraH0_2013, Verde_tension2d,BennettH0_2014}.
The low value of $H_0$ found, within the $\Lambda$CDM model,  by the Planck Collaboration since its first data release \citep{Planck13_param}, and confirmed by the latest data release \citep{Planckparameterspaper},  has attracted a lot of attention. Re-analyses  of the direct measurements of $H_0$ have been performed (\citep{EfstathiouH0_2014} including the recalibration of  distances of \cite{HumphreyH0});   physics beyond the standard model has been advocated to alleviate the tension, especially higher number of effective relativistic species, dynamical dark energy and non-zero curvature \citep{Wyman14_nu,Dvorkin14, Leistedt14, Aubourg_2015, Planck15_MGDE,  DiValentino:2016hlg}.

In some of these model extensions,
 by allowing the extra parameter to vary,  tension is  reduced but this  is mainly due to weaker constraints on $H_0$ (because of the increased number of model parameters), rather than an actual shift in the central value. In many cases, non-standard values of the extra parameter appear disfavoured by other data sets.

Recent improvements in the process of measuring $H_0$ (an 
increase in the number of SNeIa calibrated by Cepheids from 8 to 19, new parallax measurements, stronger constraints on the Hubble flow and a refined computation of distance to NGC4258 from maser data) have made possible a $2.4\%$ measurement of $H_0$: $H_0 = 73.24\pm 1.74$ ${\rm Mpc}^{-1}{\rm km/s}$ \citep{RiessH0_2016}. This  new measurement  increases the tension with respect to the latest Planck-inferred value \citep{Planck_newHFI}  to $\sim 3.4 \sigma$. This calibration of $H_0$ has been successfully tested with recent Gaia DR1 parallax measurements of cepheids  in \cite{Casertano16_Gaia}.

 Time-delay cosmography measurements of quasars which pass through strong lenses is another way to set independent constraints on $H_0$. Effort in  this direction is represented by  the H0LiCOW project \citep{Suyu_holicow}. Using three strong lenses, they find $H_0 = 71.9^{+2.4}_{-3.0}$ ${\rm Mpc}^{-1}{\rm km/s}$, within flat $\Lambda$CDM with free matter and energy density \citep{H0_holicow}. Fixing $\Omega_M= 0.32$ (motivated by the  Planck results \citep{Planckparameterspaper}), yields a value $H_0 = 72.8\pm 2.4$ ${\rm Mpc}^{-1}{\rm km/s}$. These results are in $1.7\sigma$ and $2.5\sigma$ tension with respect to the most-recent CMB inferred value, while are perfectly consistent with the local measurement of \citep{RiessH0_2016}. 

In addition, in \citep{Addison_2016}, it is shown that the value of $H_0$ depends strongly on the CMB multipole range analysed. Analysing only temperature power spectrum,  tension of 2.3$\sigma$ between the $H_0$ from $\ell < 1000$ and from $\ell \geq 1000$ is found, the former being consistent with the direct measurement of \citep{RiessH0_2016}.  
However, Ref. \citep{Planck_shifts} finds that the shifts in the cosmological parameters values inferred from low versus high multipoles are not highly improbable in a $\Lambda$CDM model (consistent with the expectations within a 10$\%$). These shifts appear because when considering only multipoles $\ell < 800$ (approximately the range explored by WMAP) the cosmological parameters are more strongly affected by the well known $\ell < 10$ power deficit.

Explanation for this tension in $H_0$ includes internal  inconsistencies in Planck data  systematics in the local determination of $H_0$  or  physics beyond the standard model. These recent results clearly motivate a detailed study of possible extensions of the  $\Lambda$CDM model and an inspection of the current cosmological data sets, checking for inconsistencies.
 
\begin{figure}[t]
\minipage{0.86\textwidth}
\begin{center}
\includegraphics[width=0.75\textwidth]{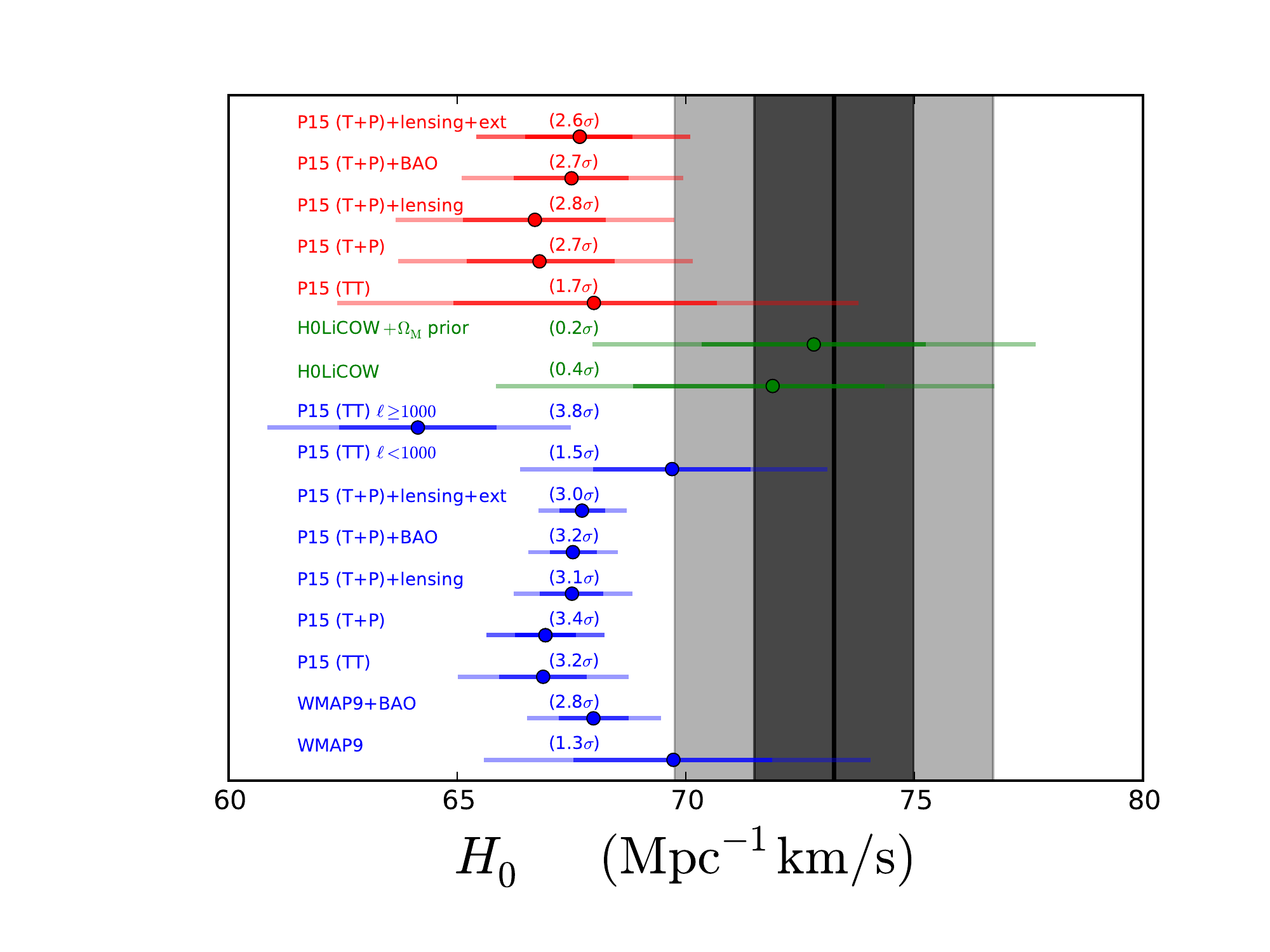}
\end{center}
\endminipage
\minipage{0.14\textwidth}
\hspace{-2cm}
\includegraphics[width=1.05\textwidth]{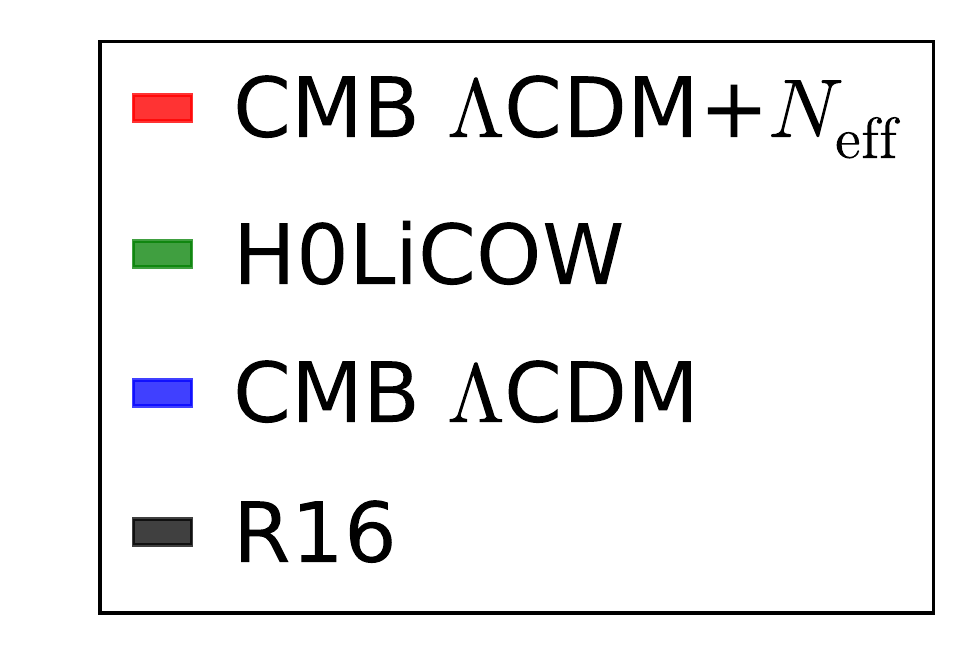}
\endminipage\hfill
\caption{\footnotesize 
Marginalised 68\% and 95\% constraints on $H_0$ from different analysis of CMB data, obtained from Planck Collaboration 2015 public chains \citep{Planckparameterspaper}, WMAP9 \citep{HinshawWMAP_13} (analysed with the same assumptions than Planck) and the results of the work of Addison et al. \citep{Addison_2016} and Bonvin et al. \cite{H0_holicow}. We show  the constraints obtained in a $\Lambda$CDM context in blue,  $\Lambda$CDM+$N_{\rm eff}$ in red, quasar time-delay cosmography results (taken from H0LiCOW project \cite{H0_holicow},  for a $\Lambda$CDM model, with and without relying on a CMB prior for $\Omega_{\rm M}$) in green and the constraints of the independent direct measurement of \citep{RiessH0_2016} in black. We report in parenthesis the tension with respect to the direct measurement. 
}
\label{fig:H0_values}
\end{figure}

In figure \ref{fig:H0_values}, we summarize the  current constraints on $H_0$ tied to the CMB and low-redshift measurements. We show  results from  the public posterior samples  provided by the Planck Collaboration 2015 \citep{Planckparameterspaper}, WMAP9 \citep{HinshawWMAP_13} (analysed with the same assumptions of Planck)\footnote{The values of $r_{\rm s}$ in WMAP's public posterior samples  were computed using the approximation of \cite{EH98}, which differs from the values computed by current Boltzmann codes and used in Planck's analysis by several percent, as pointed in the appendix B of Ref. \cite{Hamann10}. As WMAP's data have been re-analysed by the Planck Collaboration, the values reported here are all computed with the same definition.}, 
the results of the work of Addison et al. \citep{Addison_2016} and the quasar time-delay cosmography measurements of $H_0$ \citep{H0_holicow}, along with the local measurement of \cite{RiessH0_2016}.    CMB constraints are shown for two models: a standard flat $\Lambda$CDM and a model where  the effective number of relativistic species $N_{\rm eff}$ is varied in  addition to the  standard $\Lambda$CDM parameters. Of all the popular $\Lambda$CDM model extensions, this is the most promising one  to reduce the tension. 
Assuming $\Lambda$CDM, the CMB-inferred $H_0$ is consistent with the local measurement only when $\ell < 1000$ are considered (the work of Addison et al. and WMAP9). However when BAO measurements are added to WMAP9 data, the tension reappears, but at a lower level ($2.8\sigma$).

On the other hand, $r_{\rm s}$ is the standard ruler which calibrates the distance scale measurements of BAO. Since BAO measure $D_V/r_{\rm s}$ (or $D_A/r_{\rm s}$ and $Hr_{\rm s}$ in  the anisotropic analysis) the only way to constrain $r_{\rm s}$ without making assumptions about the early universe physics is combining the  BAO measurement  with  other probes of  the expansion rate (such as $H_0$,  cosmic clocks \citep{Jimenez_C}  or gravitational  lensing time delays \cite{Suyu_holicow}).  When no cosmological model is assumed,  $H_0$ and $r_{\rm s}$ are understood as  anchors of the cosmic distance ladder and the inverse cosmic distance ladder, respectively. As BAO measurements always depends on the product $H_0r_{\rm s}$ (see Equations \eqref{comovdist}, \eqref{DA} and \eqref{Dv}), when the Universe expansion history is probed by BAO, the two anchors are related by  $H_0r_{\rm s}=$ constant. 
This  was illustrated in \citep{Heavens:2014rja} and more recently in \citep{StandardQuantities}, where only weak assumptions are  made on the shape of $H(z)$, and in \citep{Cuesta:2014asa}, where the normal and inverse distance ladder are studied in the context of $\Lambda$CDM and typical extensions. 

While the model-independent measurement of $r_{\rm s}$ \citep{Heavens:2014rja} is consistent with Planck, the model-dependent value of \citep{Cuesta:2014asa} is in $2\sigma$ tension with it. Both of these measurements use $H_0\approx 73.0\pm 2.4 {\rm Mpc}^{-1}{\rm km/s}$, so, this modest tension is expected to increase with the new constraint on $H_0$.

In this paper we quantify the tension in $H_0$ and explore how it could be resolved --without invoking systematic errors in the measurements-- by studying separately changes in the early time physics and in the late time physics

We follow three avenues. Firstly, we allow the early cosmology (probed mostly by the CMB) to deviate from the standard $\Lambda$CDM assumptions, leaving unaltered the late cosmology (e.g., the expansion history at redshift below $z\sim 1000$ is given by the  $\Lambda$CDM model). Secondly, we  allow for changes in the late time cosmology, in particular  in the expansion history at $z\leq 1.3$, assuming standard early cosmology (i.e., physics is standard  until recombination, but the expansion history at late time is  allowed to be non-standard). Finally, we reconstruct in a model-independent way, the late-time expansion history without making any assumption about the early-time physics, besides assuming that the BAO scale corresponds to a standard ruler (with unknown length). By combining BAO with SNeIa and $H_0$ measurements we are able to measure the standard ruler in a model-independent way.  Comparison with the Planck-derived  determination of the sound horizon at radiation drag  allows us to assess the consistency of the two  measurements within the assumed cosmological model. 

In section \ref{sec:Data} we present the data sets used in this work and in  section \ref{sec:Methods} we  describe the methodology. We explore modifications of early-time physics  from the standard $\Lambda$CDM (leaving unaltered the late-time ones) in section \ref{sec:Earlyuniverse} while changes in the late-time cosmology  are explored in section \ref{sec:H_recon}. Here we present the findings both assuming standard early-time physics and  in a way that is independent from it. Finally we summarize the conclusions of this work in section \ref{sec:Conclusions}.

\section{Data}\label{sec:Data}

The observational data we consider are:
measurements of the Cosmic Microwave Background (CMB), Baryon Acoustic Oscillations (BAO), Type Ia Supernovae (SNeIa) and direct measurements of the Hubble constant $H_0$.

We consider the full Planck 2015 temperature (TT), polarization (EE) and the cross correlation of temperature and polarization (TE)  angular data \citep{Planckparameterspaper}, corresponding to the following likelihoods: {\it Planck}  high $\ell$ ($30\leq \ell \leq 2508$) TTTEEE for TT (high $\ell$ TT), EE and TE (high $\ell$ TEEE)  and the {\it Planck} low$\ell$ for TT, EE, TE and BB (lowP, $2\leq \ell \leq 29$). 
The Planck team \citep{Planckrobustness,Planckparameterspaper}  identifies the lowP + high $\ell$ TT as the ``recommended baseline" dataset and the  high $\ell$ polarisation (high $\ell$ TEEE) as ``preliminary", because of evidence of  low level systematics ($\sim (\mu K)^2$ in $\ell(\ell+1) C_{\ell} $).  While the level of systematic contamination does not appear to affect parameter estimation, we nevertheless present results both  excluding  and including the high $\ell$ polarisation data.
In addition, we  use the lensing reconstruction  signal for the range $40\leq L \leq 400$, which we refer to as CMB lensing. 
For some models we use the publicly available posterior samples (i.e., public chains) provided by the  Planck collaboration: $\Lambda$CDM, $\Lambda$CDM+$N_{\rm eff}$ (a base $\Lambda$CDM model with an extra parameter for the effective number of neutrino species) and  $\Lambda$CDM +$Y_{P}^{\rm BBN}$ (a base $\Lambda$CDM model with an extra parameter for the primordial Helium abundance). In addition, we use the analysis of WMAP9 data with the same assumptions of Planck, which is publicly available along with the rest of Planck data. We also  use the results of Addison et al. \citep{Addison_2016}, where the Planck's temperature power spectrum is analysed in two separate multipole ranges: $\ell < 1000$ and $\ell \geq 1000$.

We use constraints on BAO from the following galaxy surveys: Six Degree Field Galaxy Survey (6dF) \citep{Beutler11}, the Main Galaxy Sample of Data Release 7 of Sloan Digital Sky Survey (SDSS-MGS) \citep{Ross15}, the LOWZ and CMASS galaxy samples of the Baryon Oscillation Spectroscopic Survey (BOSS-LOWZ and BOSS-CMASS, respectively) \citep{Cuesta16_bao}, and the reanalysed measurements of WiggleZ \citep{Kazin14_wz}. These measurements, and their corresponding effective redshift $z_{\rm eff}$, are summarized in table \ref{tab:BAO}.  
Note that for BOSS-CMASS there is an isotropic measurement ($D_V/r_{\rm s}$) and  an anisotropic measurement ($D_A/r_{\rm s}$, $Hr_{\rm s}$), which, of course, we never combine.  
When we use the anisotropic values from BOSS-CMASS in section \ref{sec:Earlyuniverse}, we take into account that they are correlated (their correlation coefficient is 0.55). We use the covariance matrix for the measurements of WiggleZ as indicated in Ref.~\citep{Kazin14_wz}. We consider that the measurements of BOSS-CMASS and WiggleZ are independent, although the regions covered by both surveys overlap. We can do so because this overlap includes a small fraction of the BOSS-CMASS sample and the correlation is very small too (always below 4\%) \citep{Beutler16_overlap, Cuesta16_bao}, hence the constraints which come from both surveys are fairly independent.
The BOSS  collaboration also provides a BAO measurement at $z\sim 2.5$ obtained from Lyman$\alpha$ forest observed in Quasars spectra. We do not include this measurement because, as it will be clear later, our approach relies on having BAO and  SNeIa  data covering roughly the same redshift range. Considering an extra BAO point at high redshift would have increased the number of  parameters needed to describe the expansion history without improving constraints in any of the  quantities we are interested in.

The publicly available Planck 2015 posterior sampling uses a slightly different BAO data set (see Ref.~\cite{Planckparameterspaper} for details). However the small difference in the data set does not drive any significant effect in the parameter constraints.

\begin{table}
\small
\begin{center}
\begin{tabular}{|cccc|}
\hline
Survey	& $z_{\rm eff}$	& Parameter	& Measurement\\
\hline
6dF \citep{Beutler11}	& 0.106	& $r_{\rm s}/D_V$	& $0.327\pm 0.015$ \\
SDSS-MGS \citep{Ross15}        & 0.15  & $D_V/r_{\rm s}$       & $4.47\pm 0.16$\\
BOSS-LOWZ \citep{Cuesta16_bao}	& 0.32	& $D_V/r_{\rm s}$	& $8.59\pm 0.15$\\
WiggleZ \citep{Kazin14_wz}  & 0.44 & $D_V/r_{\rm s}$ & $11.6\pm 0.6$ \\
BOSS-CMASS \citep{Cuesta16_bao}	& 0.57	& $D_V/r_{\rm s}$	& $13.79\pm 0.14$\\
BOSS-CMASS \citep{Cuesta16_bao}	& 0.57	& $D_A/r_{\rm s}$	& $9.52\pm 0.14$\\
BOSS-CMASS \citep{Cuesta16_bao}	& 0.57	& $Hr_{\rm s}$	& $14750\pm 540$\\
WiggleZ \citep{Kazin14_wz}  & 0.6 & $D_V/r_{\rm s}$ & $15.0\pm 0.7$ \\
WiggleZ \citep{Kazin14_wz}  & 0.73 & $D_V/r_{\rm s}$ & $16.9\pm 0.6$ \\
\hline
\end{tabular}
\end{center}
\caption{\footnotesize
BAO data measurements included in our analysis, specifying the survey that obtained each measurement and the corresponding effective redshift $z_{\rm eff}$. In the case where we change the late time cosmology, we use the isotropic measurements. We take into account the correlation between the anisotropic measurements of BOSS-CMASS and among the values from WiggleZ.
}
\label{tab:BAO}
\end{table}
 
For SNeIa cosmological observations, we use the SDSS-II/SNLS3 Joint Light-curve Analysis (JLA) data compilation \citep{Betoule14_jla}. This catalog contains 740 spectroscopically confirmed SNeIa obtained from low redshift samples ($z<0.1$), all three seasons of the Sky Digital Sky Survey II (SDSS-II) ($0.05<z<0.4$) and the three years of the SuperNovae Legacy Survey (SNLS) ($0.2<z<1$) together with nine additional SNeIa at high redshift from HST ($0.8<z<1.3$). We use the compressed form of the JLA likelihood (Appendix E of Ref.~\citep{Betoule14_jla}).

Finally, we use the distance recalibrated direct measurement of $H_0$ from \citep{RiessH0_2016}, which is $H_0 = 73.24 \pm 1.74\  {\rm Mpc^{-1}km/s}$.

\section{Methods} \label{sec:Methods}

We use the public Boltzmann code CLASS \cite{Lesgourgues:2011re, Blas:2011rf} and the Monte Carlo public code Monte Python \citep{Audren13_mp} to 
analyse the 
CMB data sets discussed in  section \ref{sec:Data} when for the selected model 
there are no  posterior samples officially provided by   the Planck collaboration. We  modify the codes to include the parametrized extra dark radiation, $\Delta N_{\rm eff}$ and the effective parameters to describe its behaviour, $c_{\rm s}^2$ and $c_{\rm vis}^2$ (section \ref{sec:Earlyuniverse}) additional parameters  to the Planck ``base"  model \footnote{The Planck ``base" model is a flat, power law power spectrum $\Lambda$CDM model with three neutrino species, with total mass 0.06eV)} .
We adopt uniform priors for all the parameters, except for $\Delta N_{\rm eff}$, for which we sample $\Delta N_{\rm eff}^2$ (see section \ref{sec:Earlyuniverse}). We only set a lower limit in the sampling range for $A_s, n_s, \tau$ and $\Delta N_{\rm eff}$ (0.0 in all cases but for $\tau$, which is $0.04$). The prior in $\tau$ has virtually no effect on the reported constraints and is justified by observations of the Gunn--Peterson effect, see e.g., Ref. \cite{Caruana14}. 
 Our sampling method of choice when CMB data are involved is the   Metropolis Hastings algorithm; we    run sixteen Monte Carlo Markov Chains (MCMC) for each ensemble of data sets until the fundamental parameters reach a convergence parameter $R-1 < 0.03$, according to the Gelman-Rubin criterion \citep{Gelman92}.
 
When   interpreting the low redshift  probes of the  expansion history (see section \ref{sec:H_recon}),  we use a different methodology.  We aim to reconstruct $H(z)$ (the main observable related with the expansion of the Universe) in the most model-independent way possible, but still requiring a smooth expansion history. For this reason  the Hubble function is   expressed as piece-wise natural cubic splines in the redshift range $0 \leq z \leq 1.3$.  We  specify the spline function $H^{\rm recon}(z)$ by the values it takes  at $N$ ``knots" in redshift. These  values  uniquely define the piecewise cubic spline once we ask for continuity of $H^{\rm recon}(z)$ and   its first and second derivatives at the knots, and two boundary conditions. We require the second derivative to vanish at the exterior knots.
Thus, our free parameters are the values of $H^{\rm recon}(z_{\rm knot})$, where $z_{\rm knot}$ are the redshifts correspondent to the ``knots". We also consider cases in which we vary the sound horizon at radiation drag and the curvature of the Universe (via $\Omega_{\rm k}$). The location of the knots is arbitrary and we place them at  $z = \left[0, 0.2, 0.57, 0.8, 1.3 \right]$  to match the BAO data  constraining power and  encompass the SNeIa redshift range. When  SNeIa are not included, we limit the fit of $H^{\rm recon}(z)$ to the range $0\leq z \leq 0.8$ (and vary one less parameter).  
 
We set uniform priors for all the parameters, with  limits which are  never explored by the MCMC. We use the public \texttt{emcee} code \citep{emcee}, which implements the Affine Invariant Markov Ensemble sampler as sampling method \citep{Goodman_mcmc} to fit the splines to the cosmological measurements discussed in the previous section. To obtain the likelihood of each position in the parameter space, we integrate the correspondent $H^{\rm recon}(z)$ to compute $D_V(z)$ and luminosity distance  $D_L(z)$ and calculate the $\chi^2$ for  BAO and SNeIa, respectively. In addition, we fit $H^{\rm recon}(z=0)$ to the direct measurement of $H_0$. We run 500 walkers for 10000 steps each and remove the first 400 steps from each walker (as burn-in phase), as this interval corresponds to several autocorrelation times.

In section \ref{sec:H0-rs}, we quantify the tension between the  different joint  constraints on the plane $H_0$-$r_{\rm s}$ following \citep{Verde_tension2d}. This method is based on the evidence ratio of the product of the distributions with respect to the --ideal , and ad hoc-- case when  the maxima of the posteriors  coincide (maintaining shape and size).

Then, if we call $P_{A}$ to the posterior of the experiment $A$ and $\mathcal{E}$ to the `unnormalized' evidence, and with a bar we refer to the shifted case,
\begin{equation}
\mathcal{T} = \frac{\bar{\mathcal{E}}\mid_{{\rm max} A = {\rm max} B}}{\mathcal{E}} = \frac{\int \bar{P}_A\bar{P}_B dx}{\int P_A P_B dx} .
\end{equation}
$\mathcal{T}$ is the degree of tension and can be interpreted in the modified Jeffrey's scale. The odds for the null hypothesis (i.e. both posteriors are fully consistent) are 1 : $\mathcal{T}$. 

\section{Modifying early Universe physics: effect on $H_0$ and $r_{\rm s}$}\label{sec:Earlyuniverse}
It is well known that there are two promising ways to alter early cosmology  so that the tension between CMB-inferred value  and measured value of $H_0$ is reduced. These are changing the early time expansion history  and changing the details of recombination. 

\begin{figure}[h]
 \begin{center}
 \minipage{0.5\textwidth}
    \includegraphics[width=\textwidth]{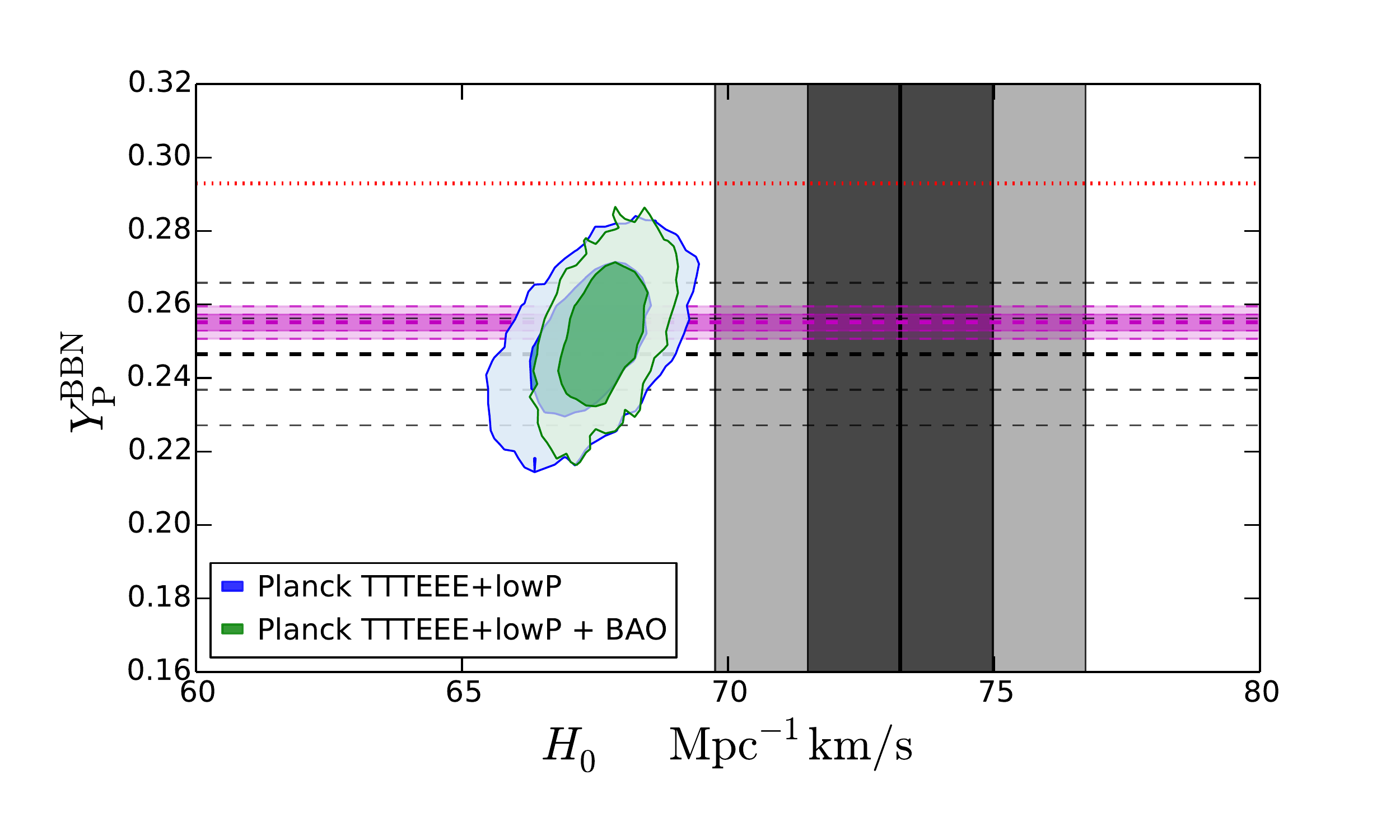}
    \endminipage
    \minipage{0.5\textwidth}
    	\includegraphics[width=\textwidth]{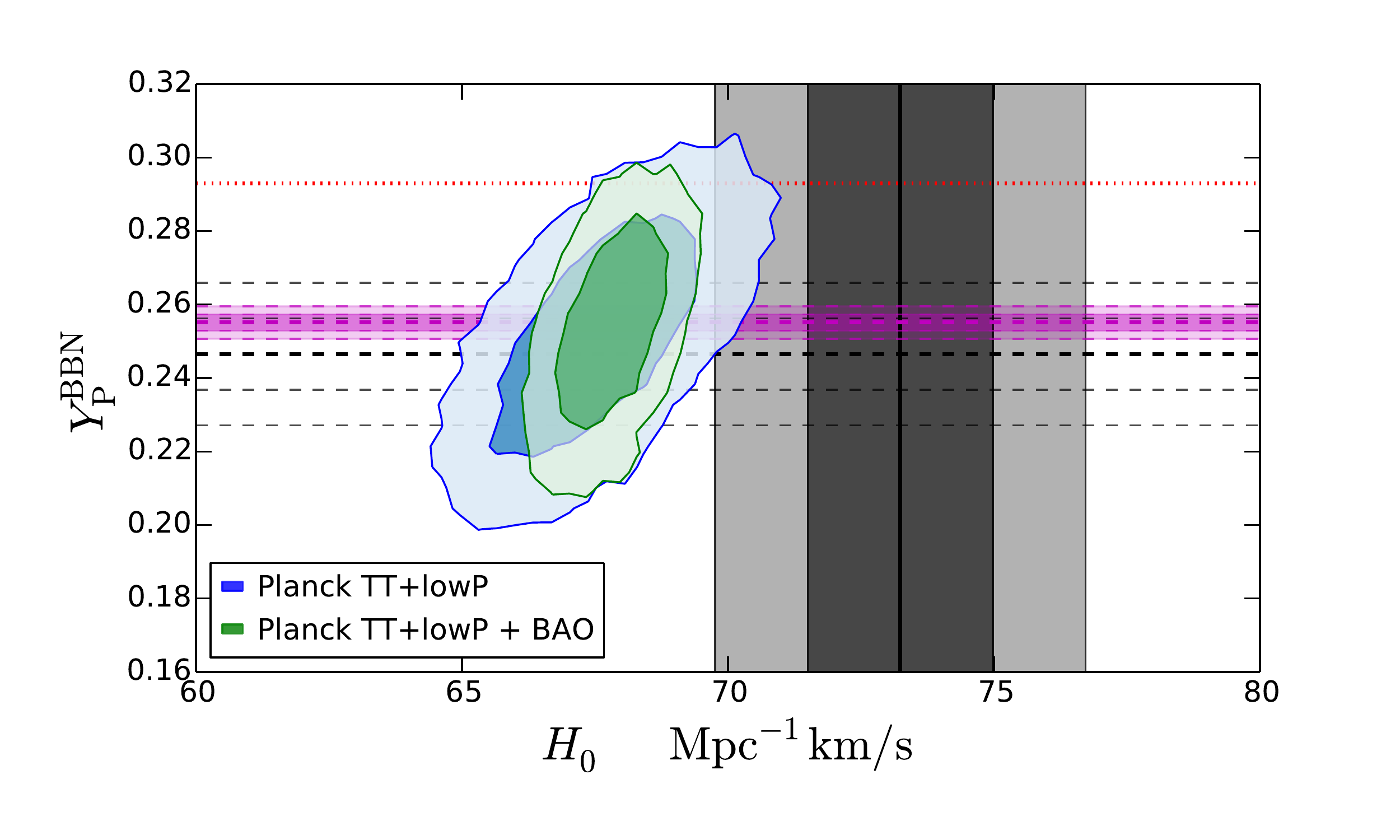}
    	\endminipage\hfill
     \caption{$68$\% and $95$\% confidence  joint constraints in the $H_0$-$Y_{\rm P}^{\rm BBN}$ parameter space for Planck 2015 using temperature and polarization power spectra (left) and without include high $\ell$ polarization data (right).
    The vertical bands correspond to the local $H_0$ measurement \cite{RiessH0_2016}.  The horizontal black dashed lines  correspond to the measurement  (mean and 1 and 2 $\sigma$) of the primordial abundance of \cite{Aver:2013wba}, and in magenta of \citep{Izotov_Yp_2014}, both from chemical abundances in metal-poor HII regions. The red dotted horizontal line is the 2 $\sigma $ upper limit of the recent measurement of initial Solar helium abundance  of \cite{Serenelli:2010fk}.}
    \label{fig:Yp}
  \end{center}
\end{figure}

Changes in the  details of nucleosynthesis can be captured by changes in the primordial  Helium mass fraction,
 parametrised by $Y_{\rm P}^{\rm BBN}$. In the standard analyses,  since the  process of standard big bang
nucleosynthesis (BBN) can be accurately modelled and gives a
predicted relation between $Y_{\rm P}^{\rm BBN}$, the photon-baryon ratio, and the
expansion rate, the value of $Y_{\rm P}^{\rm BBN}$ is computed consistently with BBN for every model sampled, but one can also relax any BBN prior and let $Y_{\rm P}^{\rm BBN}$ vary freely, which has an influence on the recombination history and affects CMB anisotropies
mainly through the redshift of last scattering and the diffusion
damping scale. The effect of this extra degree of freedom on the inferred value of $H_0$ can be seen in figure \ref{fig:Yp} (obtained using the publicly released Planck team's MCMC),  the local $H_0$ measurement \cite{RiessH0_2016},  and the  measurements  (mean and 1 and 2 $\sigma$) of the primordial abundance of \cite{Aver:2013wba}  and \cite{Izotov_Yp_2014} (which is less conservative) from chemical abundances in metal-poor HII regions and the conservative $95\%$ upper limit of the measured initial Solar helium abundance of \citep{Serenelli:2010fk}.

Even varying $Y_{\rm P}^{\rm BBN}$ without a BBN prior, the joint $H_0$-$Y_{\rm P}^{\rm BBN}$ constraints are in a $\sim 2.7\sigma$ disagreement (when using lowP and high $\ell$ TTTEEE) with the new measurement of $H_0$ \citep{RiessH0_2016}. If high $\ell$ polarization data is not included, the tension is reduced because of the larger error bars. However, the constraints from Planck  are not in agreement with both $H_0$ and primordial abundance measurements at the same time, even considering the more conservative measurement of \citep{Aver:2013wba}.

\begin{figure}[h]
 \begin{center}
 \minipage{0.5\textwidth}
    \includegraphics[width=\textwidth]{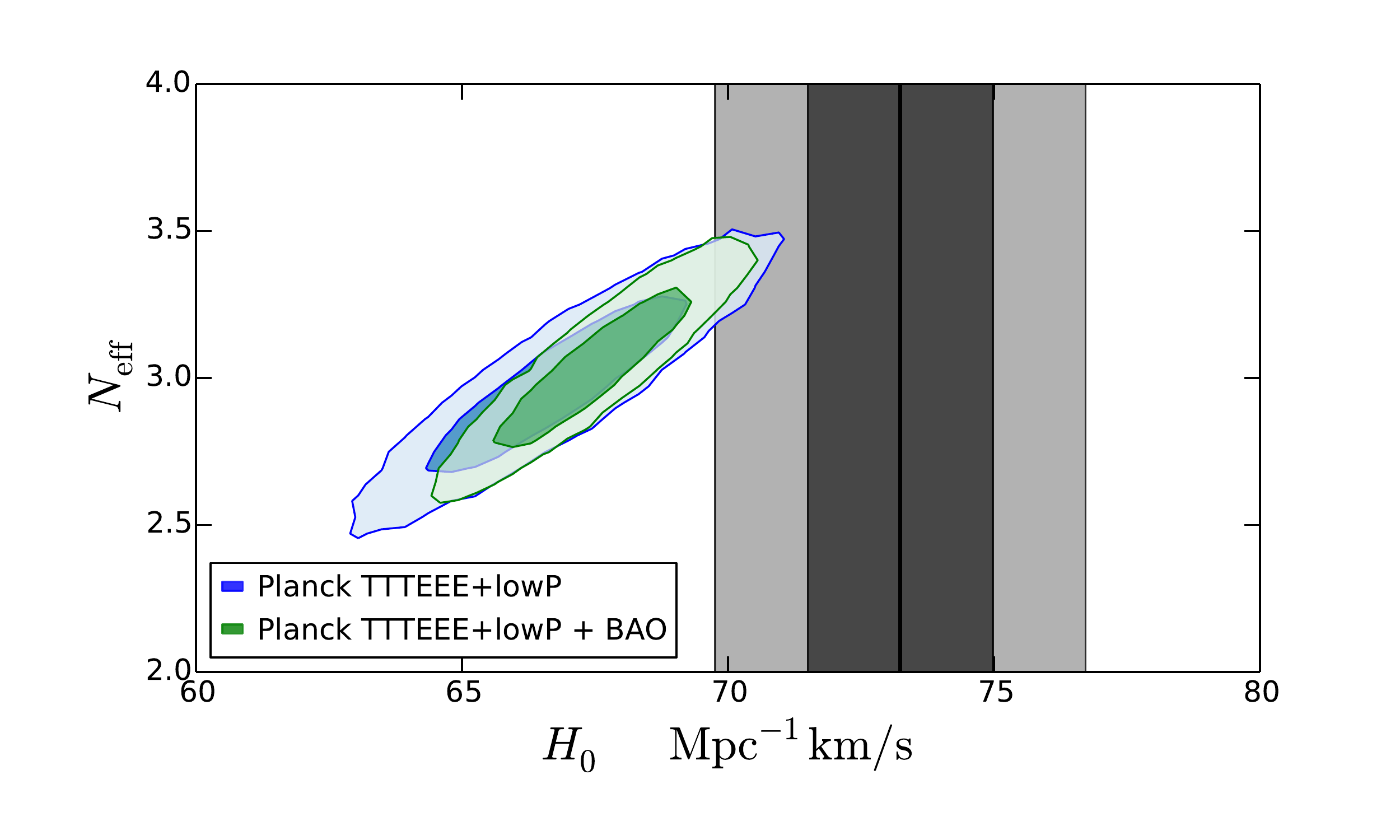}
 \endminipage 
 \minipage{0.5\textwidth}
 	\includegraphics[width=\textwidth]{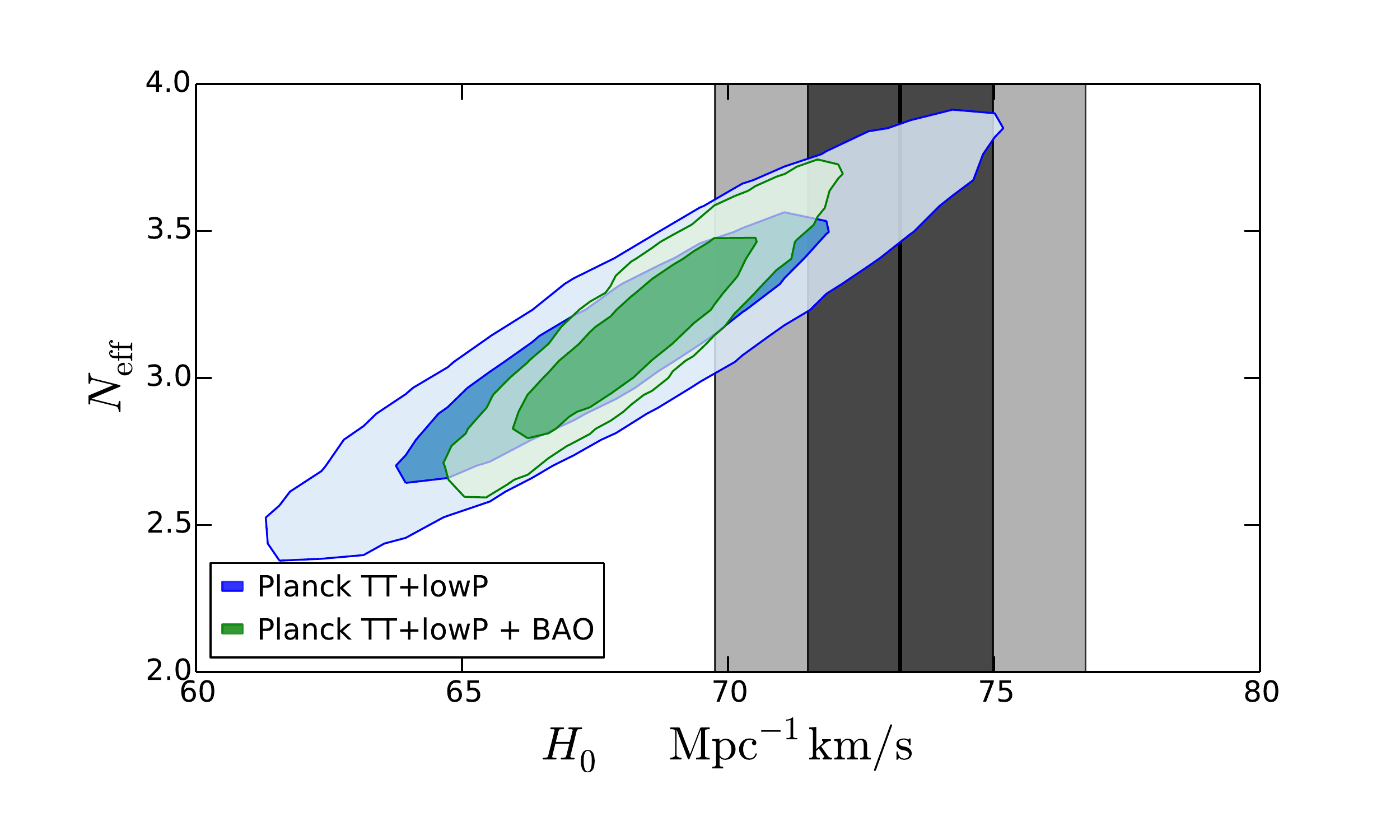}
 \endminipage\hfill
    \caption{Confidence regions (68\% and 95\%) of the  joint constraints in the $H_0$-$N_{\rm eff}$ parameter space for Planck 2015 data (blue) and  Planck 2015 + BAO data (green) using  full temperature and polarization power spectra (left) and without including high $\ell$ polarization data (right). Here all species behave like neutrinos when perturbations are concerned. The vertical bands correspond to the local $H_0$ measurement \cite{RiessH0_2016}. }
    \label{fig:Neff_planck}
  \end{center}
\end{figure}

Changes on the early time expansion history are  usually enclosed in the parameter $N_{\rm eff}$: the effective number of relativistic species. For three standard neutrinos  $N_{\rm eff}=3.046$\footnote{The number of (active) neutrinos species is 3, the small correction accounts for the fact that  the neutrino decoupling epoch was immediately followed by $e^+e^-$ annihilation.}
 \cite{Mangano:2005cc}.  In fact light neutrinos are  relativistic at decoupling time and they behave like radiation:  changing  $N_{\rm eff}$ changes the composition of the  energy density,  changing therefore the  early expansion history.  This has been called ``dark radiation"  but it can mimic several other physical effects see e.g., \cite{Ackermanetal, Abazajian:2012ys, Weinberg:2013kea, Kelso:2013paa,  Mastache:2013iha, DiBari:2013dna, Archidiacono:2013fha,  Boehm:2012gr, Hasenkamp:2012ii,   Anchordoqui:2014dpa, SolagurenBeascoa:2012cz, GonzalezGarcia:2012yq}.  For example a model  such as  the one proposed in  \cite{Weinberg:2013kea}  of a thermalized massless boson, has a $\Delta N_{\rm eff}$ between $\sim 0.57$ and $0.39$ depending on the decoupling temperature \cite{Planckparameterspaper}. 
 
 If we define $\Delta N_{\rm eff}$ as $N_{\rm eff}-3.04$, it is well known that a $\Delta N_{\rm eff}>0$ would increase the CMB-inferred $H_0$ value, bringing it closer to the locally measured one. This can be appreciated in   Fig.~\ref{fig:Neff_planck}, where we show the results of Planck 2015 for a model where   $N_{\rm eff}$ is an additional free parameter and the extra radiation behaves like neutrinos.  In the $H_0$-$N_{\rm eff}$ parameter space  we show the joint    68\% and 95\% confidence regions  for Planck 2015 data (blue) and  Planck 2015 + BAO data (green) obtained from the Planck team's public chains, both using polarization and temperature power spectra (left) or just temperature power spectrum and lowP (right). The vertical bands correspond to the local $H_0$ measurement \cite{RiessH0_2016}.

 A high value of $N_{\rm eff}$ ($\Delta N_{\rm eff}\sim 0.4$) would alleviate the tension in $H_0$ and still  be allowed by the Planck lowP and high $\ell$ temperature power spectra and BAO data as pointed out  in \citep{RiessH0_2016}.  The ``preliminary" high $\ell$ polarization data, disfavours such large $\Delta N_{\rm eff}$ (at $\sim 2\sigma$ level), as polarization constrains strongly the effective number of relativistic species. 
  
  This is however not the full story. State-of-the-art CMB data have enough statistical power to measure not just the effect of this $N_{\rm eff}$ on the expansion history but also on the perturbations. Neutrino density/pressure perturbations, bulk velocity and anisotropic stress are additional sources for the gravitational potential via the Einstein equations (see e.g., \cite{Bashinsky:2003tk, Hou:2011ec, Lesgourgues:1519137}).  The effect on the perturbations is described by the effective parameters sound speed and viscosity $c_s^2$, $c_{\rm vis}^2$ \cite{Hu:1998tk, Hu:1998kj, Trotta:2004ty, Smith:2011es}. Neutrinos have $\{c_s^2, c_{\rm vis}^2\}=\{1/3,1/3\}$, but other values describe other physics, for example a perfect relativistic fluid will have $\{1/3,0\}$ and a scalar field oscillating in a quartic potential  $\{1,0\}$. Different values of $c_s^2$ and $c_{\rm vis}^2$  would describe other dark radiation candidates. This parametrisation is considered flexible enough for providing a good approximation to several alternatives to the standard case of free-streaming particles e.g., \cite{Cyr-Racine:2013jua, Oldengott:2014qra}.

 Recent analyses have shown that if all  $N_{\rm eff}$ species have the same effective parameters  $c_s^2$, $c_{\rm vis}^2$, Planck data constraints are tight \cite{Audren:2014lsa, Planckparameterspaper}: $c^2_{s}=0.3240\pm0.0060$, $c^2_{\rm vis}=0.327\pm0.037$ (with fixed $N_{\rm eff}=3.046$; Planck 2015). Moreover, the $N_{\rm eff}$ constraints are not significantly affected compared to the standard case: $N_{\rm eff}=3.22^{+0.32}_{-0.37}$ (\citep{Audren:2014lsa}) against $N_{\rm eff}=3.13\pm0.31$ (Planck 2015) at 68\% confidence level, both using CMB temperature data and CMB lensing. We update the results of \citep{Audren:2014lsa} by using Planck 2015 power spectra (lowP + high $\ell$ TTTEEE) instead of Planck 2013. Results can be seen in table \ref{tab:all_Neff}.
 Using state-of-the-art observations, the constraints on are even tighter than in \citep{Audren:2014lsa} and are driven by the high $\ell$ polarisation data.  As the polarisation analysis is ``preliminary" these results should be considered preliminary too. In all the cases studied, there is no significant shift in the central value of $H_0$. 
There is no evidence for the main component of the relativistic species behaving differently from a standard neutrino, and this extension does not  alleviate the tension in $H_0$ significantly (tension is reduced only because extending the model results in a  slightly larger uncertainty in  $H_0$).

\begin{table}[h]
\scriptsize
\begin{center}
\begin{tabular}{|c|c|c|c|c|}
\hline
 & $N_{\rm eff}$ & $c^2_s$ & $c^2_{\rm vis}$ & $H_0$ \\
\hline
 lowP+TTTEEE & $2.96\pm 0.23$	& $0.324\pm 0.006$	& $0.33 \pm 0.04$ &	$67.2\pm 1.9$\\
lowP +TTTEEE+ lensing &	$2.91\pm 0.21$	& $0.325\pm 0.006$	& $0.33\pm 0.04$	&	$67.0\pm 1.8$	\\
lowP +TTTEEE+lensing+BAO & $2.94\pm 0.18$	& $0.325	\pm 0.006$	&$0.33 \pm 0.04$	& $67.2\pm 1.3$		\\
\hline
\end{tabular}
\end{center}
\caption{Marginalised mean and 68\% confidence level errors for the parameters of interest for the different combinations of data.}
\label{tab:all_Neff}
\end{table}

\begin{figure}[t]
\includegraphics[width=\textwidth]{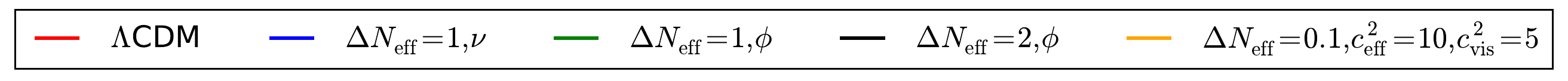}
\minipage{0.5\textwidth}
\begin{center}
\includegraphics[width=\textwidth]{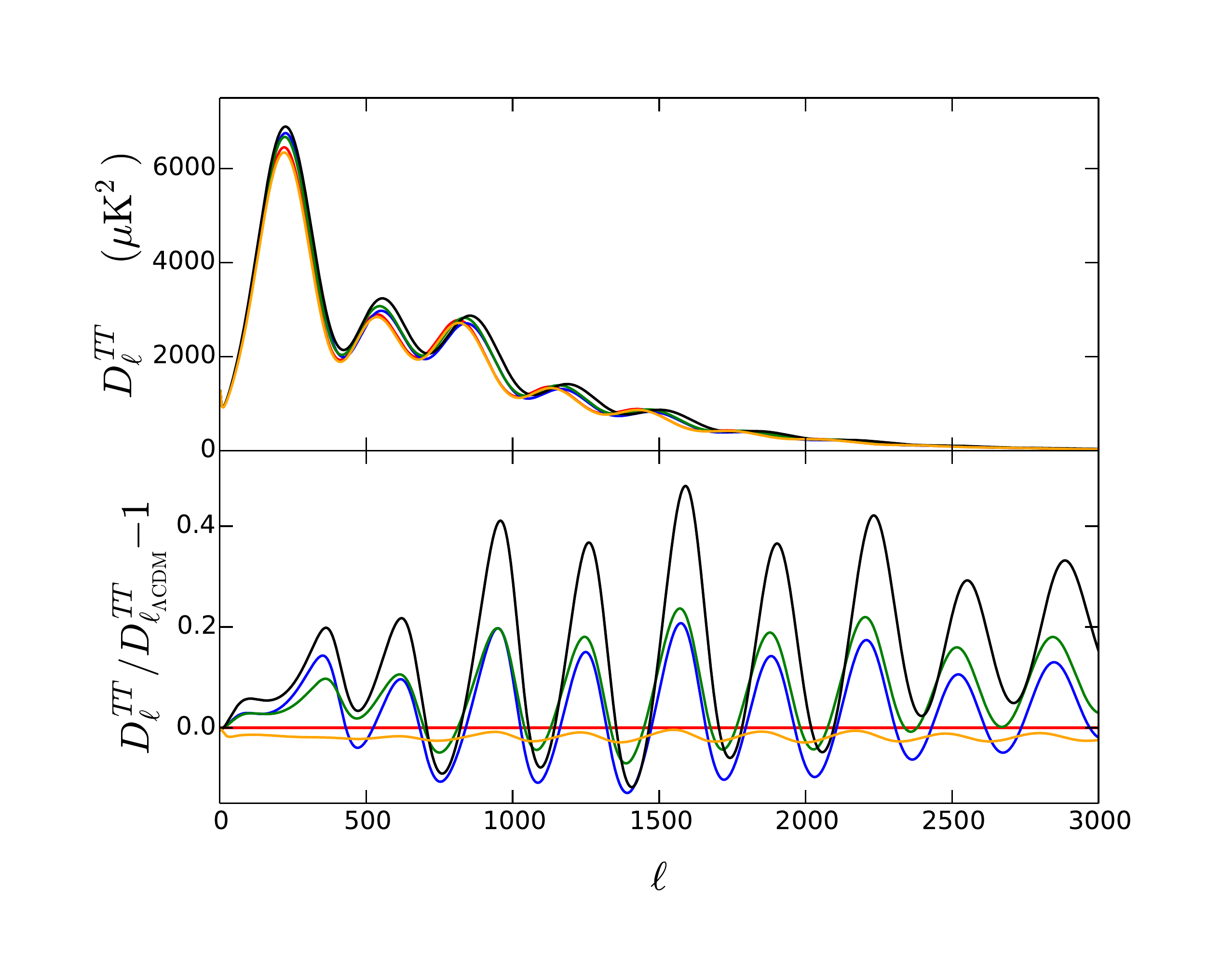}
\end{center}
\endminipage
\minipage{0.5\textwidth}
\begin{center}
\includegraphics[width=\textwidth]{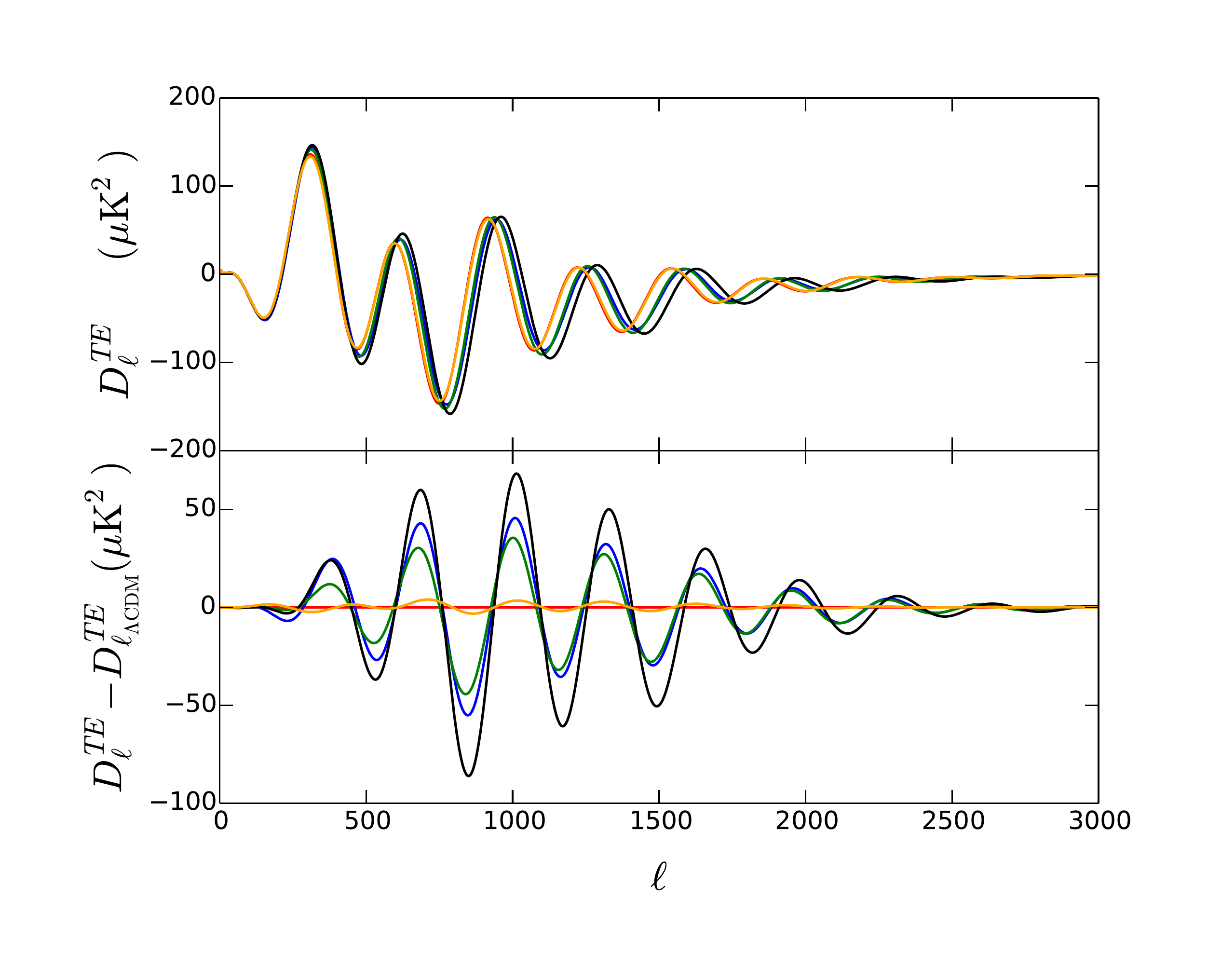}
\end{center}
\endminipage\hfill
\vspace{-0.5cm}
\minipage{0.5\textwidth}
\begin{center}
\includegraphics[width=\textwidth]{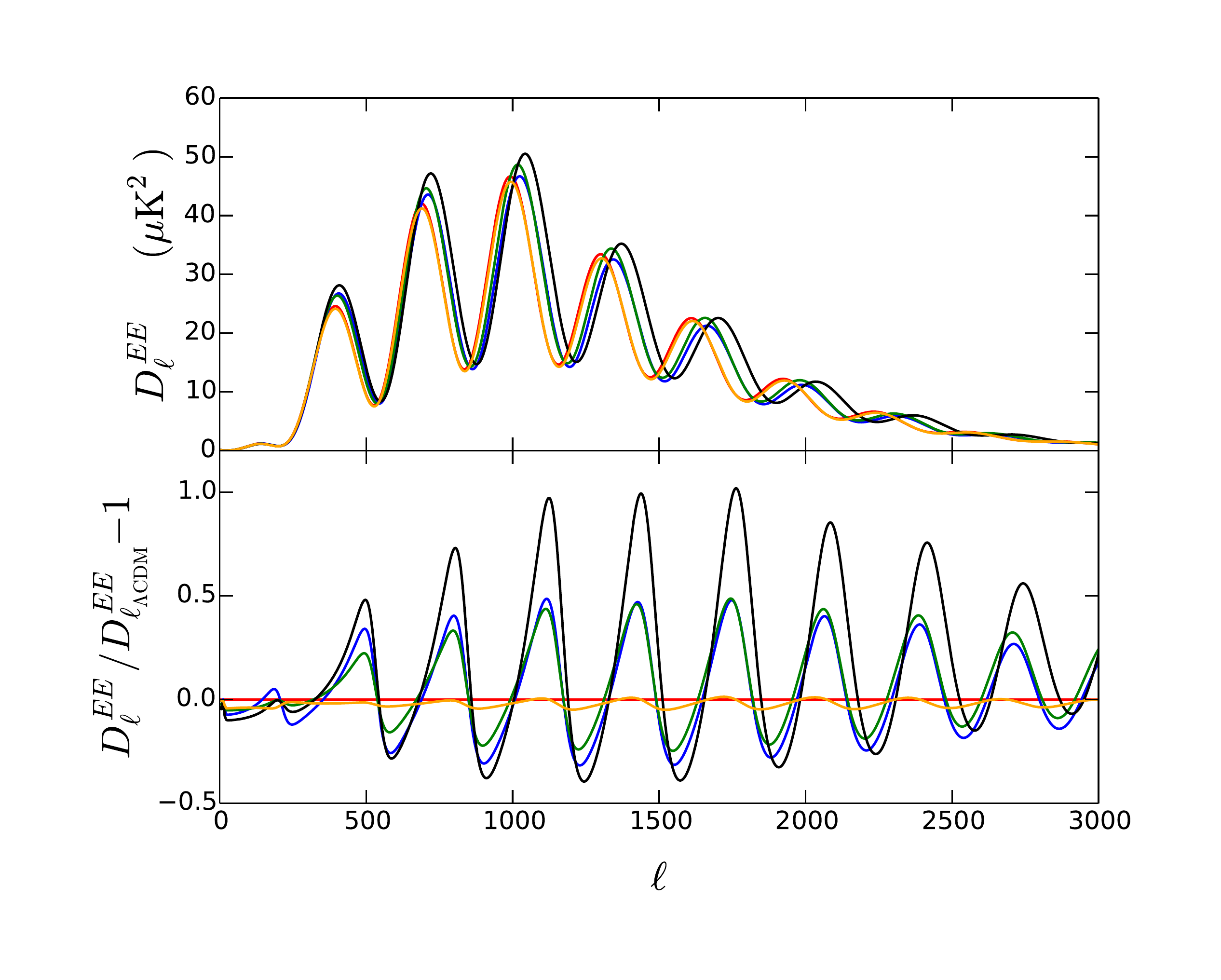}
\end{center}
\endminipage\hfill
\caption{
CMB temperature ({\it left}), temperature and polarization cross correlation ({\it right}) and polarization ({\it bottom}) power spectra predictions for $\Lambda$CDM (red) and the following extensions: one more neutrino (blue), one scalar field (green), two scalar fields (black) and a illustrative case with extreme (non physical) values of $c^2_s$ and $c^2_{\rm vis}$ with $\Delta N_{\rm eff}=0.1$ (orange). 
}
\label{fig:ceffcvis_test}
\end{figure}

The fact that  when leaving $c_s^2$, $c_{\rm vis}^2$ as free parameters in the analysis one recovers  tight constraints consistent with $\{1/3,1/3\}$  and $N_{\rm eff}=3.04$ is a good confirmation of the existence of a cosmic neutrino background. However,  this does not exclude the possibility of the existence of extra  relativistic species (i.e., dark radiation) with different behaviour than neutrinos. Their presence could be masked in the analysis by the dominant component, the cosmic neutrino background. Thus next, we shall assume that there are three neutrino families in the Universe (i.e., that the 3.04 effective species have  $\{c_s^2$, $c_{\rm vis}^2\}$=$\{1/3,1/3\}$) and that any extra dark radiation $\Delta N_{\rm eff}$ component has  free effective sound and viscosity parameters. We have modified the publicly available CLASS code \cite{Lesgourgues:2011re, Blas:2011rf} to implement this.  
In figure \ref{fig:ceffcvis_test} we show the qualitative effects on the CMB power spectra of the parameters describing this extra dark radiation component. 

\begin{figure}[h]
\minipage{0.5\textwidth}
\begin{center}
\includegraphics[width=\textwidth]{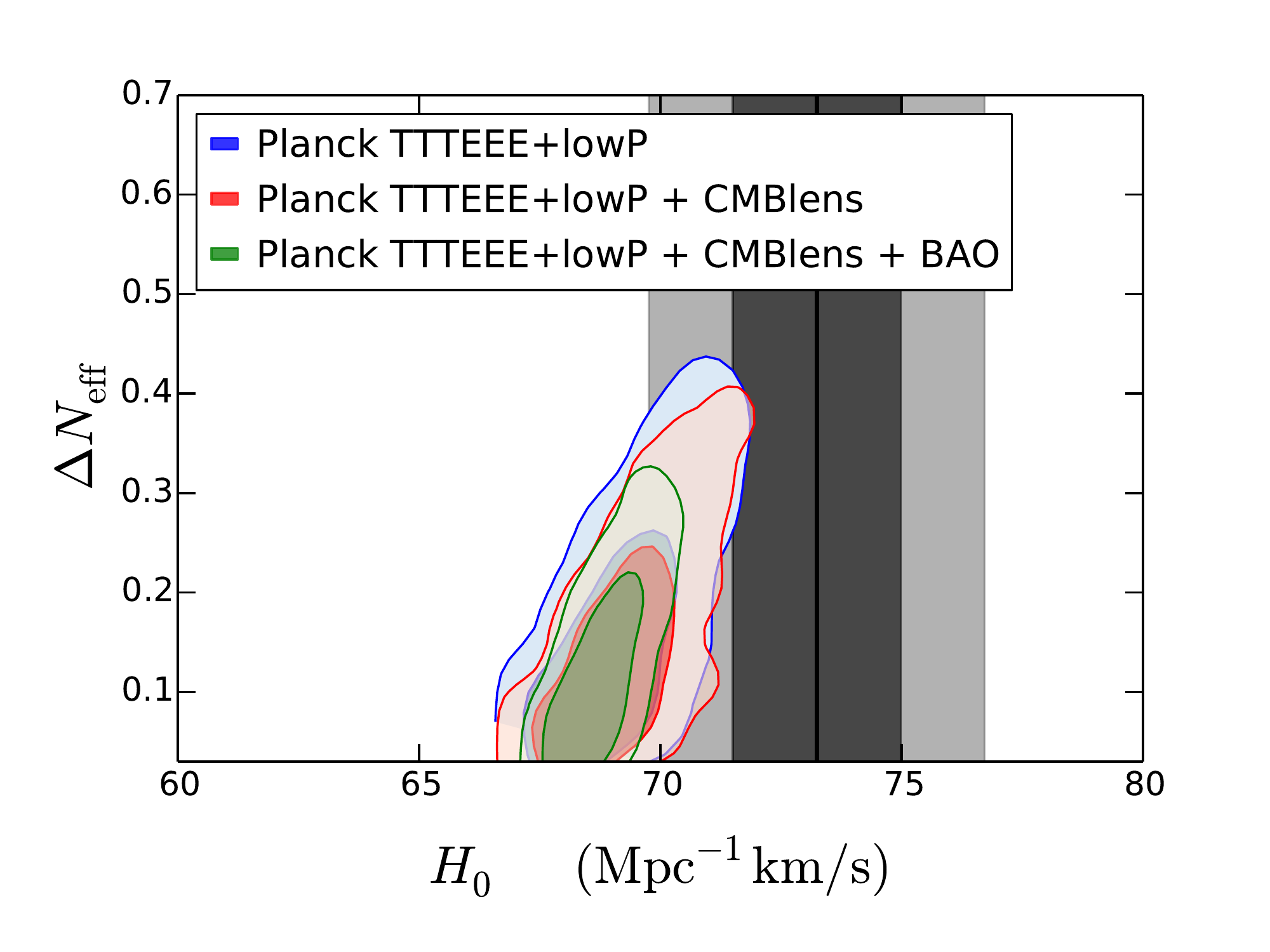}
\end{center}
\endminipage
\minipage{0.5\textwidth}
\begin{center}
\includegraphics[width=\textwidth]{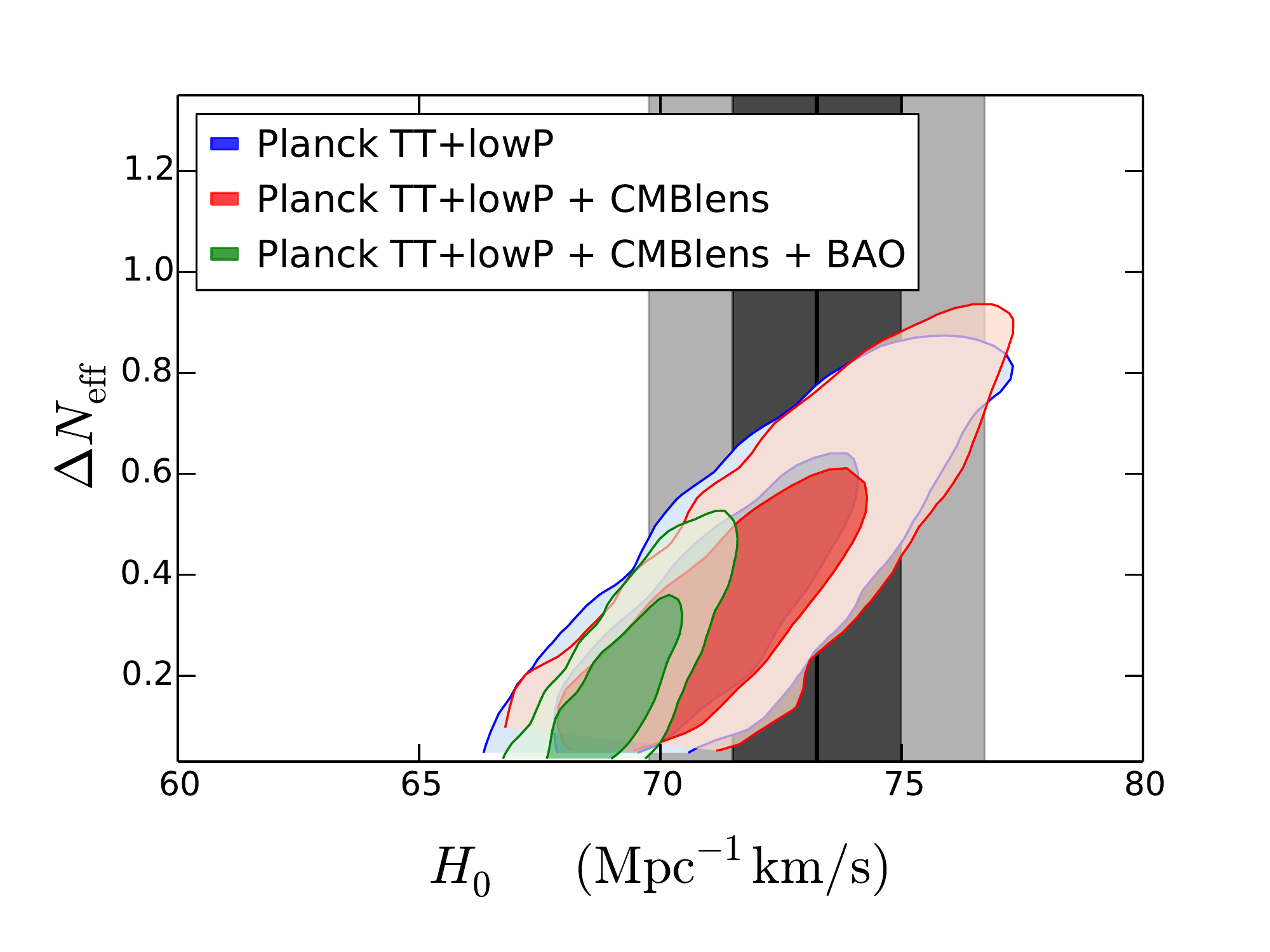}
\end{center}
\endminipage\hfill
\caption{
Marginalized 68$\%$ and $95\%$ confidence level constraints in the $\Delta N_{\rm eff}$-$H_0$ plane.  \textit{Left}: Planck  data including full  temperature and polarization power spectra. \textit{Right}: Excluding high $\ell$ polarisation. We report results using Planck 2015 power spectra (blue), adding CMB lensing (red) and adding also BAO (green). The vertical black bands correspond to the local $H_0$ measurement \citep{RiessH0_2016}. Note the change in the scale of the y axis in each plot.}
\label{fig:Neff_ceffcvis}
\end{figure}

Given that for $\Delta N_{\rm eff}\sim 0$ there is almost no difference in the likelihood for different values of $c_s^2$ and $c_{\rm vis}^2$, the MCMC tend to be stuck in that zone of the parameter space \footnote{For the parameters $c_{\rm vis}^2$ and $c_s^2$  we limit the sampling to the range $[0.0,1.1]$. That is because values higher than one are not physical and this way we also optimize the performance of the analysis, since we do not explore the parameter space where these values are allowed to be very  large in the region close to  $\Delta N_{\rm eff} = 0$.}. To prevent it, we sample $\Delta N_{\rm eff}^2$ instead. We find that, once 3.046 standard neutrinos are fixed, the presence of extra relativistic species, even giving freedom to the behaviour  of their  perturbations, is not favoured by the data. The constraints on $c_{\rm vis}$, $c_s$ are very weak because of the strong upper limit on $\Delta N_{\rm eff}$.
 The results are summarized in table \ref{tab:Delta_Neff}, with the upper limits of $\Delta N_{\rm eff}$ at 95\% of confidence level. We also report the values for the $\Omega_{\rm M}-\sigma_8$ combination, as this is very similar to what  galaxy clusters or weak gravitational lensing constrain. While $\Delta N_{\rm eff}$ and $\sigma_8$ are slightly correlated (an increase of 0.5 in  $\Delta N_{\rm eff}$  increases  $\sigma_8$ by 0.03) the  constraint  on the $\Omega_{\rm M}-\sigma_8$ combination  do not depend on the limits on $\Delta N_{\rm eff}$. The joint constraints on $\Delta N_{\rm eff}$ and $H_0$ are shown in figure \ref{fig:Neff_ceffcvis}.

\begin{table}[h]
\scriptsize
\begin{center}
\begin{tabular}{|c|c|c|c|c|c|}
\hline
 & $\Delta N_{\rm eff}$ & $c^2_s$ & $c^2_{\rm vis}$ & $H_0$ & $\left(\Omega_{\rm M}/0.3\right)^{0.5}\sigma_8$\\
\hline
lowP+ TTTEEE & $<0.36$	& $0.25^{+0.07}_{-0.15}$	& $0.12^{+0.58}_{-0.11}$ &	$68.9^{+1.1}_{-0.9}$ 	& $0.85^{+0.02}_{-0.02}$\\
lowP+ TTTEEE+CMB lensing &	$<0.34$	& $0.24^{+0.10}_{-0.13}$	& $0.49\pm 0.33$	&	$68.9^{+1.2}_{-0.9}$	& $0.83^{+0.02}_{-0.01}$	\\
lowP+ TTTEEE+CMB lensing+BAO & $<0.28$	& $0.26	^{+0.09}_{-0.16}$	&$0.28 ^{+0.45}_{-0.26}$	& $68.7^{+0.6}_{-0.7}$ &	$0.84^{+0.01}_{-0.02}$	\\
lowP + TT & $<0.76$	& $0.25^{+0.08}_{-0.10}$	& $0.84^{+0.20}_{-0.51}$ &	$70.6^{+2.6}_{-2.0}$	& $0.84^{+0.03}_{-0.04}$ \\
lowP+ TT+CMB lensing &	$<0.77$	& $0.27^{+0.07}_{-0.11}$	& $0.81^{+0.19}_{-0.50}$	&	$71.3^{+1.9}_{-2.2}$	&  $0.82^{+0.02}_{-0.01}$\\
lowP+ TT+CMB lensing+BAO & $<0.44$	& $0.29	^{+0.20}_{-0.16}$	&$0.9^{+0.1}_{-0.7}$	& $69.0^{+0.9}_{-0.8}$	& 	$0.84^{+0.01}_{-0.02}$\\
\hline
\end{tabular}
\end{center}
\caption{Marginalised constraints for the parameters of interest for the different combinations of data. We report the the upper limit for $\Delta N_{\rm eff}$ (95\% confidence level) and the highest posterior density intervals for the rest of the parameters.}
\label{tab:Delta_Neff}
\end{table}

 Should the low-level systematic present in the polarisation data be found to be sub-dominant in the published error-budget,
this finding implies that there is not much room for an extra component in the early universe whose density  scales with the expansion  like radiation but whose perturbations  have the freedom to behave like a perfect fluid, a neutrino, a scalar field or anything in between.   This offers a useful confirmation of  one of the key standard assumptions on which the standard cosmological model is built.   Also in this more general model, of the CMB data, it is  the high-$\ell$ polarisation  what disfavor high values of $H_0$. On the other hand, the freedom on the nature of the extra relativistic species produces a small shift in $H_0$ towards higher values and, when high $\ell$ polarization data of the CMB are not included, our constraints are fully compatible with the direct measurement (right panel of figure \ref{fig:Neff_ceffcvis}). However, when BAO data are included, the constraint on $\Delta N_{\rm eff}$ is tighter, because the degeneracy with $\Omega_{\rm M}$.

\section{Changing late-time cosmology}\label{sec:H_recon}
The CMB is sensitive to both late and early cosmology. When fitting the CMB power spectrum  
simultaneous assumptions about the early and late cosmology must be made, with the implication that the physics of both epochs are entwined in the resulting constraints. Then, it is difficult to determine what   physics beyond $\Lambda$CDM would be the responsible of possible deviations from the model. 
Exploring non-standard  late cosmology evolution, possibly in a minimally-parametric or model independent way  is in general complicated if CMB constraints are to be included. 
There is however a way to analyse CMB data so that  it is sensitive only to  early cosmology as shown in  \cite{Vonlanthen:2010cd, Audren:2013nwa, Audren:2012wb}:  the resulting constraints do not depend on late-time physics and can  therefore be included when analysing late-time data in a model-independent way.  The latest CMB Planck data were analysed in this way in \cite{edepaper}, where a variety of models of the early Universe are studied.  Here we use their results --obtained  assuming standard early-time physics (i.e. a flat Universe composed of baryons, radiation,  standard neutrinos, cold dark matter and dark energy in the form of  cosmological constant  from deep in the radiation era down to recombination)-- in the form of a constraint on the sound horizon at radiation drag:  $r_{\rm s}^{\rm early}=147.00\pm 0.34$ Mpc.

To reconstruct $H(z)$, we use BAO and SNeIa data along with the measured value of $H_0$ and $r_s^{\rm early}$. The only assumptions made are that the expansion history of the Universe is smooth and continuous, that the spatial section of the Universe is flat, that SNeIa form an homogeneous group such as they can be used as standard candles and that the sound horizon at radiation drag, $r_{\rm s}$, is a standard ruler which calibrates the cosmic distance scale given by BAO observations (\cite{Cuesta:2014asa, Heavens:2014rja,StandardQuantities}). We also consider the case in which the assumption about the geometry of the Universe is relaxed. With this minimal assumptions, $H_0$ and $r_{\rm s}$ are treated only as the calibration of the cosmic distance ladder and they are related by $H_0r_{\rm s}=$ constant.

Without any assumption about the geometry of the Universe, the comoving distance $\chi$ is related to the Hubble parameter by
\begin{equation}
\chi (z) = \frac{c}{H_0\sqrt{\lvert\Omega_{\rm k}\rvert}}\mathcal{S}_{\rm k}\left(\sqrt{\lvert\Omega_{\rm k}\rvert}H_0 \int_0^z \frac{dz^\prime}{H(z^\prime)}\right) \,,
\label{comovdist}
\end{equation}
where $\mathcal{S}_{\rm k}(x) = \sinh(x)$, $x$ or $\sin(x)$ for $\Omega_{\rm k} > 0 $, $=0$ or $<0$, respectively. Then, the angular diameter distance and the luminosity distance are
\begin{equation}
D_A(z) = \frac{\chi(z)}{(1+z)}\,, \quad\quad\quad\quad D_l(z) = (1+z)\chi(z)\,.
\label{DA}
\end{equation} 

BAO observations provide measurements of $D_V$, which is related to $H(z)$ by
 \begin{equation}
 D_V(z) = \left[cz\left(1+z\right)^2 D_A(z)^2H(z)^{-1}\right]^{1/3} = \left[cz\left(\int_0^{z}\frac{c dz^\prime}{H(z^\prime)} \right)^2H(z)^{-1}\right]^{1/3} \,,
 \label{Dv}
 \end{equation}

where the last identity is true only when flatness is assumed. 
While it is customary to parametrise dark energy properties via the equation of state parameter $w(z)$, it should be evident that the observable quantity is $H(z)$. Afterwards, to convert $H(z)$ into $w(z)$ a model for dark energy must be assumed as well as a value for $\Omega_m$:
\begin{equation}
H(z)=(1+z)^{3/2}\sqrt{\Omega_m+\Omega_{\rm DE}\exp\left[3\int^z_0\frac{w(z')}{1+z'}dz'\right]}\,,
\end{equation}
here $\Omega_{\rm DE}$ is the density parameter for the dark energy and flatness is assumed.

For our  model-independent reconstruction of the late-time expansion history where the Hubble parameter  is expressed  in piece-wise natural cubic splines, $H(z)$ is specified by  the values it takes  at $N$ ``knots" in redshift; $N=4$ for BAO only analysis and $N=5$ when SNeIA are included. This parametrisation allows for smooth and relatively generic expansion histories. As  indicated in \citep{Betoule14_jla} for the compressed data set of supernovae, we allow an offset in the absolute magnitude compared with the standard value, $\Delta M$, treated as a free (nuisance) parameter. 

\begin{table}
\scriptsize
\begin{center}
\makebox[\textwidth]{
\begin{tabular}{|c|c|c|c|c|c|c|c|}
\hline
Data sets	& $H(z=0)$	& $H(z=0.2)$	&$H(\!z=0.57)$ & $H(z=0.8)$ & $H(z=1.3)$ & $r_{\rm s}$& $\Delta M$ \\
\hline
$H_0$+BAO+$r_{\rm s}^{\rm early}$ & $72.3\pm 1.7$ & $72.9 \pm 1.9$ & $96.4\pm 2.5$ & $102.5\pm 14.0 $ & -- & -- & --  \\
$H_0$+SN & $73.2\pm 1.8$ & $81.0 \pm 2.5$ & $99.3\pm 4.4$ & $107.0\pm 9.0 $ & $161.9\pm 73.1$ & -- &$0.10\pm 0.06$ \\
$H_0$+BAO +$r_{\rm s}^{\rm early}$+SN & $69.4\pm 1.0$ & $75.5 \pm 1.2$ & $94.0\pm 1.8$ & $101.2\pm 6.2 $ & $150.1\pm 62.9$ & --&  $-0.03\pm 0.03$  \\
$H_0$+BAO(*)+SN & $73.1\pm 1.8$ & $80.6\pm 2.4$ & $101.5\pm 3.8$ & $109.3\pm 7.6$ & $143.7\pm 59.7$ & \!$136.8\pm 4.0$\! & $0.10\pm 0.06$ \\
\hline
$H_0$+BAO+$r_{\rm s}^{\rm early}$+SN($\circ$) & $69.6_{-1.3}^{+1.1}$ & $75.6\pm 1.2$ & $94.0 \pm 4.1$ & $101.1 \pm 11.2$ & $147.1\pm 89.3$ & -- & $-0.03\pm 0.04$\\
$H_0$+BAO(*)+SN($\circ$) & $73.4^{+1.5}_{-2.0}$ & $83.0\pm 3.0$ & $111.9 \pm 8.9$ & $130.0\pm 17.8$ & $237.9\pm 123.4 $ & \!$133.0 \pm 4.7$\! & $0.10\pm 0.06$ \\
BAO+$r_{\rm s}^{\rm early}$+SN($\circ$) & $66.3_{-1.7}^{+1.7}$ & $75.1\pm 1.1$ & $101.2\pm 5.7$ & $118.1 \pm 13.9$ & $215.9\pm 112.0$ & -- & $-0.11\pm 0.05$ \\
\hline
\end{tabular}}
\end{center}
\caption{\footnotesize
Marginalized mean and 68\% confidence regions  for the parameters included in the reconstruction for each of the combinations of data sets we consider. When reporting asymmetric errors, we report the highest posterior density value for $H_0$. The last column corresponds to $\Delta M$, the offset in the absolute magnitude compared with the standard value.  We report it here to show that the  supernovae absolute magnitude is not  significantly  shifted away from its value determined internally  by  external  the data. the ``*" symbol indicate that no CMB-derived $r_{\rm s}$ prior  is used. The symbol ``$\circ$" indicates that $\Omega_{\rm k}$ is left as a free parameter.}
\label{tab:H_recon}
\end{table}

We summarize the results of our analysis (mean and  68\% confidence intervals) in table \ref{tab:H_recon}. In the following figures (figures \ref{fig:H_recon_bao}-\ref{fig:H_recon_BAO+SN}) we show the best fit of our reconstruction of $H(z)$ (black line), the 68\% confidence region obtained by plotting the curves corresponding to 500 points of the chain randomly selected  from the 68\%  with highest likelihood (red)
and the Hubble parameter corresponding to the $\Lambda$CDM prediction using the  best fit values of Planck 2015 with lensing \citep{Planckparameterspaper} (blue). The dashed blue  line is the $\Lambda$CDM prediction using   $H_0=73.0$ ${\rm Mpc}^{-1}$km/s instead of the  Planck 2015 inferred value. In the plots, data  are shown in green and the predictions of the observables using $H^{\rm recon}(z)$ in black (bestfit) and red (68\% region). In this case we show  ratios with respect $\Lambda$CDM prediction for clarity. The vertical grey lines mark the position of the ``knots". 

\begin{figure}[h]
\minipage{0.5\textwidth}
\begin{center}
\includegraphics[width=\textwidth]{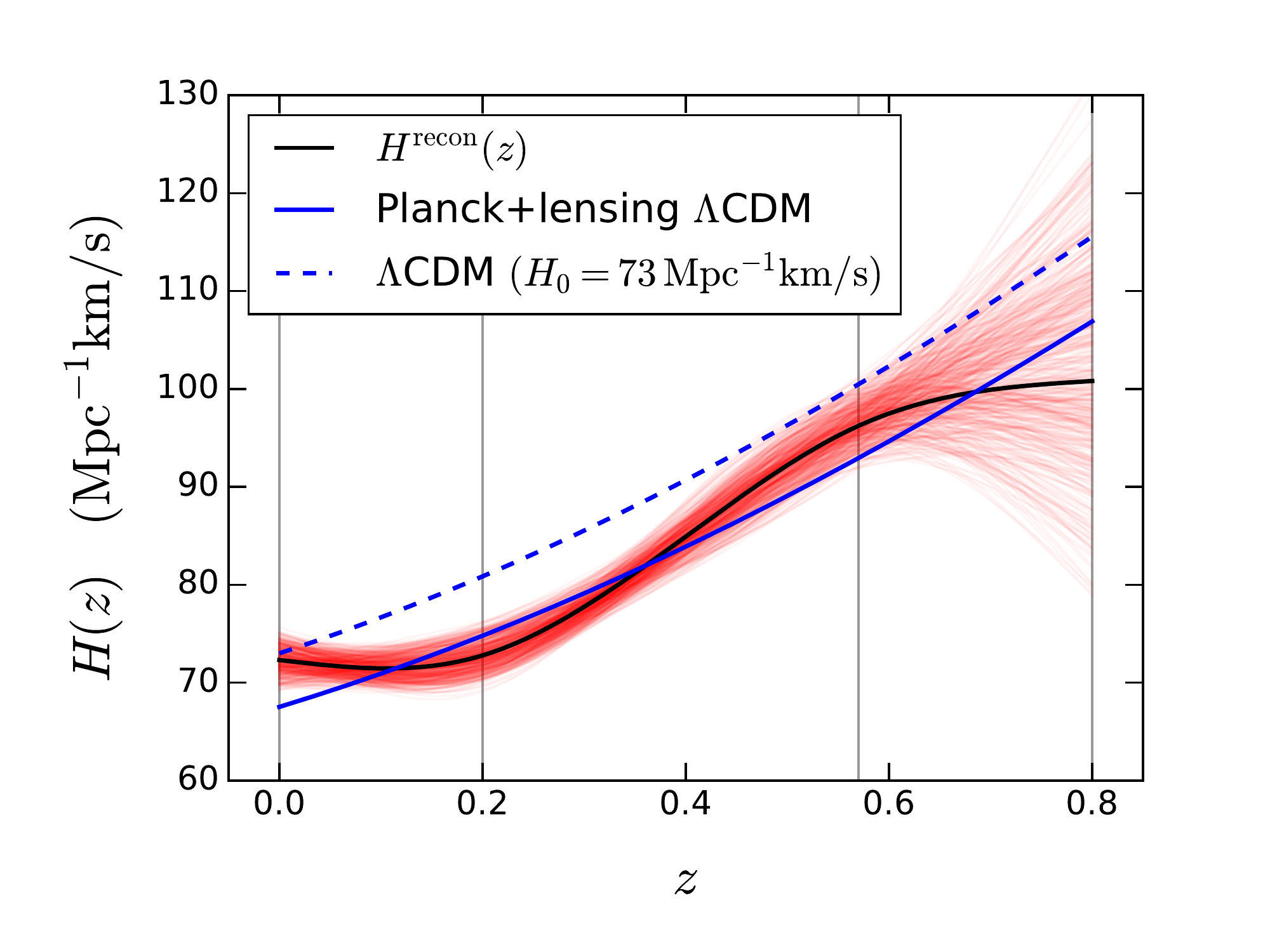}
\end{center}
\endminipage
\minipage{0.5\textwidth}
\begin{center}
\includegraphics[width=\textwidth]{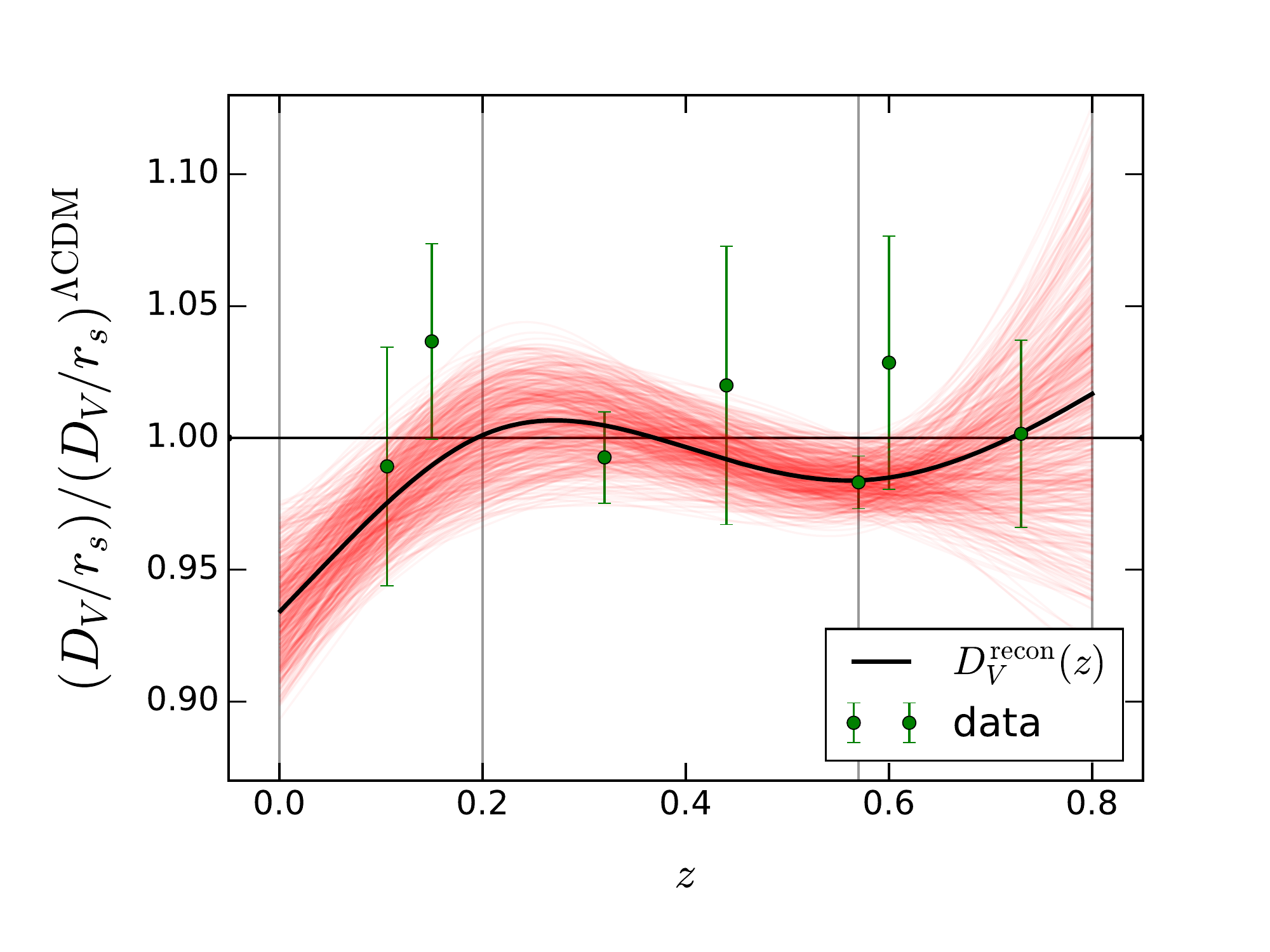}
\end{center}
\endminipage\hfill
\caption{\footnotesize
\textit{Left}: Results of the reconstruction of $H(z)$ using the direct measurement of $H_0$ of \citep{RiessH0_2016} and the BAO data set. 
\textit{Right}: BAO data included in the reconstruction. We plot $(D_V/r_{\rm s})r_{\rm s}^{\rm early}$ data points and the $D_V$ obtained with the corresponding $H^{\rm recon}(z)$.
}
\label{fig:H_recon_bao}
\end{figure}

 Figure \ref{fig:H_recon_bao} shows the reconstruction  of $H^{\rm recon}(z)$ and  $D_V(z)$ for the  analysis of BAO and  $H_0$ \citep{RiessH0_2016} data with the $r^{\rm early}_{\rm s}$ prior. 
 While in a $\Lambda$CDM model $H(z)$ is monotonically increasing with redshift, here $H^{\rm recon}(z)$ is almost constant in the range $0<z<0.2$ to match the local $H_0$ determination to  the distance measurements which are ``anchored" by $r_{\rm s}^{\rm early}$, predicting a sharper acceleration at low redshift. Given that the lowest redshift sampled by the  BAO data is $z=0.106$, $H_0$ is determined mostly by the direct measurement of \citep{RiessH0_2016}. Using the formalism described in section \ref{sec:Methods}, we can quantify the significance of this feature with respect to the Planck 2015  $\Lambda$CDM $H(z)$ distribution.
 We consider various redshifts ($z=0$, 0.2 and 0.57 which we select to  coincide with the knots)  yielding a  multivariate distribution. 
 With this choice the odds of obtaining the same results are 1:49.  Although the results would be different depending on the chosen redshifts, we consider that this choice is representative and the results will not vary qualitatively with another reasonable choice. This applies for all the cases studied here.

In  figure \ref{fig:H_recon_SN} we show the results for the reconstruction of $H(z)$ using  SNeIa and $H_0$. We show  $H^{\rm recon}(z)$ and the distance modulus. The redshift sampling of SNIa data is much denser than BAO: this constrains the shape of $H^{\rm recon}(z)$, but not the normalisation (as the analysis marginalises over the supernovae absolute magnitude) which is  anchored at $H_0 \sim 73$ ${\rm Mpc}^{-1}$km/s by the $H_0$ measurement. The reconstructed shape  is  very close to the  $\Lambda$CDM until the  data  sampling is sparser ($z\gtrsim 0.6$) and errors grow. The odds of SNeIa reconstructed $H(z)$ shape compared to  the one obtained using only BAO, $H_0$ and $r_{\rm s}^{\rm early}$ are 1:52.

\begin{figure}[h]
\minipage{0.5\textwidth}
\begin{center}
\includegraphics[width=\textwidth]{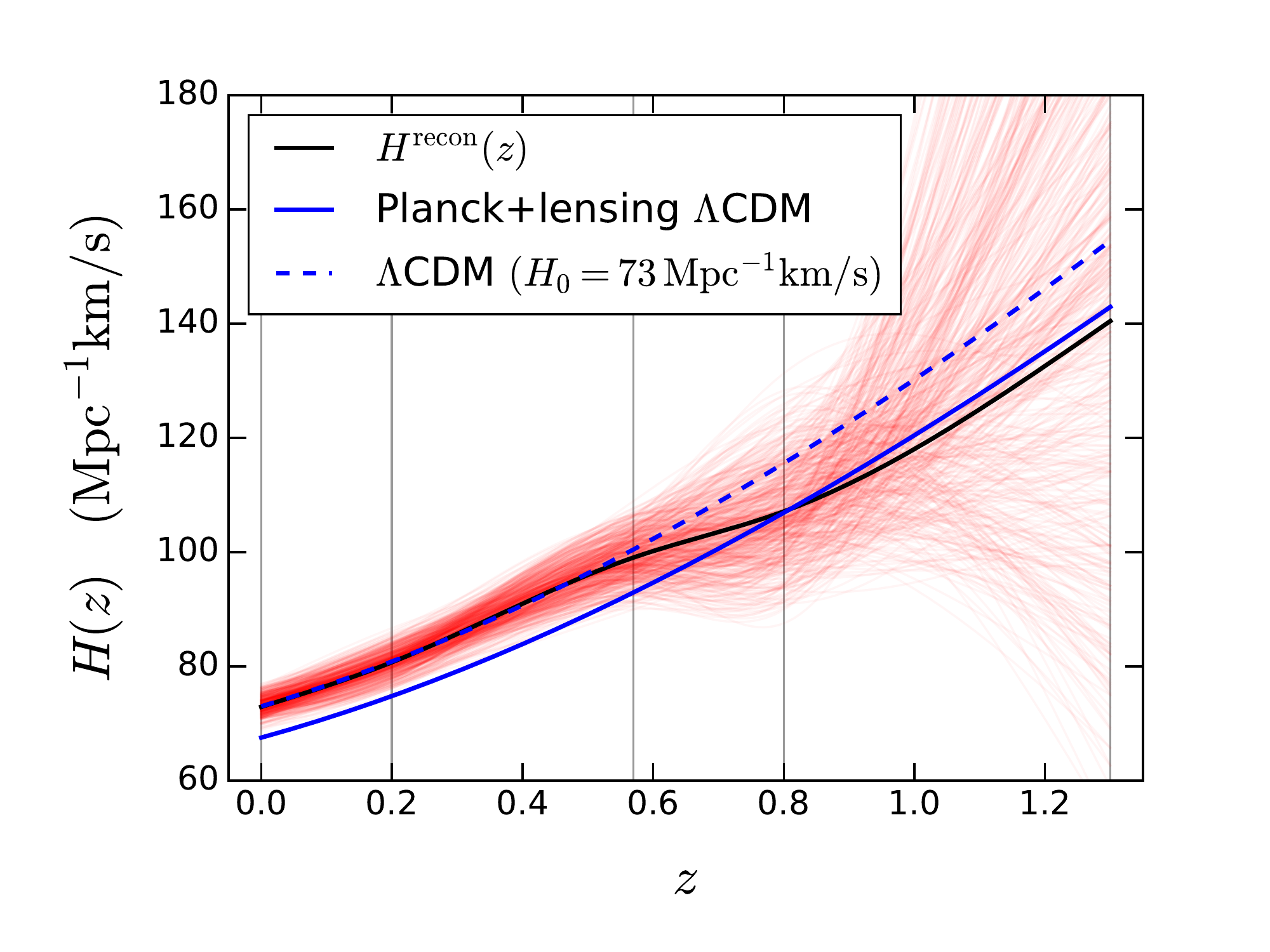}
\end{center}
\endminipage
\minipage{0.5\textwidth}
\begin{center}
\includegraphics[width=\textwidth]{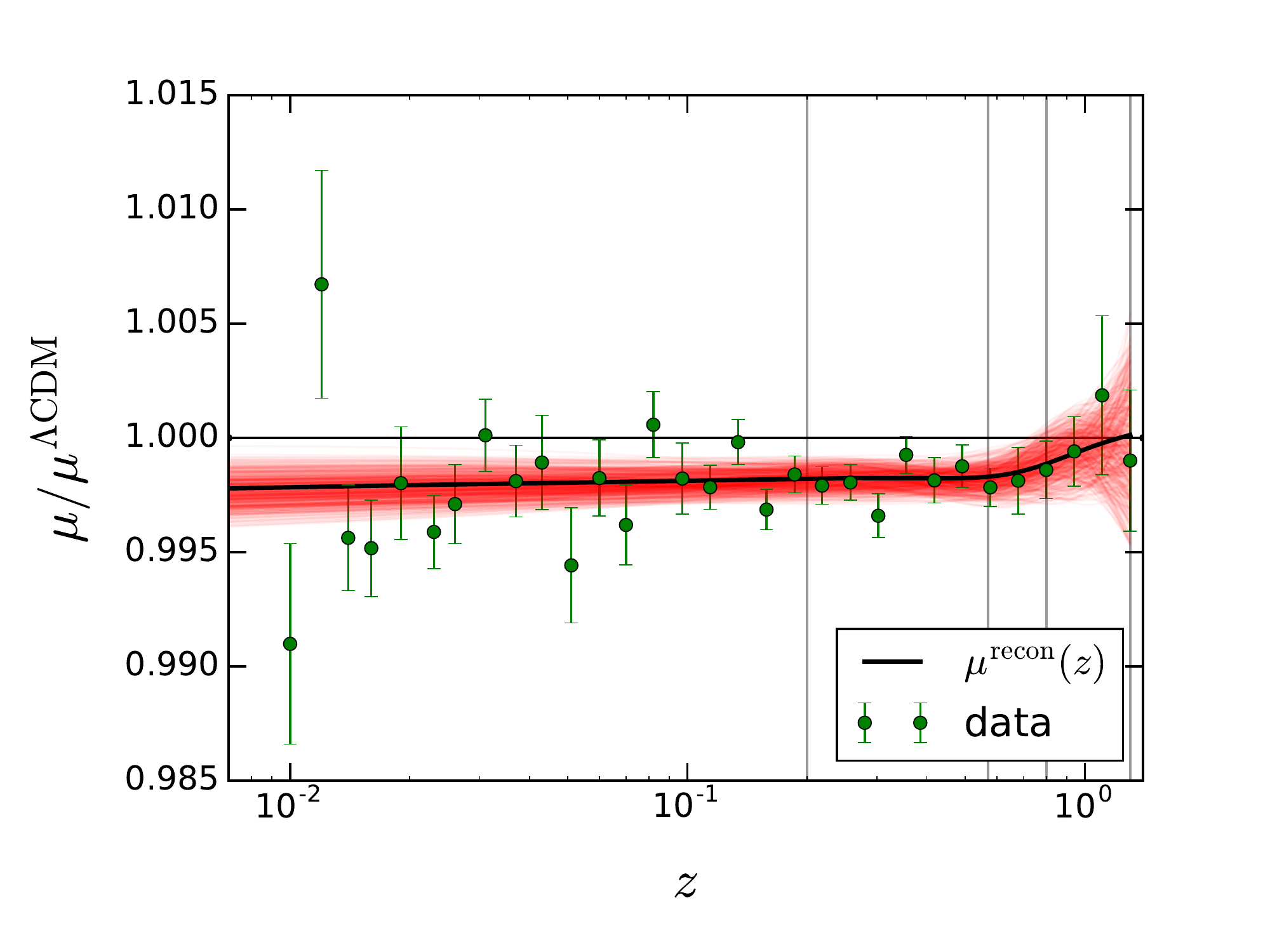}
\end{center}
\endminipage\hfill
\caption{\footnotesize
\textit{Left}: Results of the reconstruction of $H(z)$ using the direct measurement of $H_0$ of \citep{RiessH0_2016} and the SNeIa data set. 
\textit{Right}: SNeIa data included in the reconstruction (green). We plot the distance modulus, $\mu$, obtained with  the corresponding $H^{\rm recon}(z)$. The plotted errorbars correspond to the square root of the diagonal of the covariance matrix (we account correctly for the actual correlation among bins  in the analysis).
}
\label{fig:H_recon_SN}
\end{figure}

Finally, when using the combination  $H_0$, SNeIa and BAO with our $r^{\rm early}_{\rm s}$ prior (figure \ref{fig:H_recon_BAO+SN}), SNeIa observations constrain the shape of $H^{\rm recon}(z)$ but the normalization tries to fit $H_0$ and BAO (via $r_{\rm s}^{\rm early}$) at the same time. The $H_0$ measurement has a 2.4\% error, but the $r_{\rm s}^{\rm early}$ determination a 0.23\% error: the statistical power of the BAO normalisation  shifts the recovered $H^{\rm recon}(z=0)$   to lower  values   compared to the local determination (and closer to the Planck--inferred value under a $\Lambda$CDM model). Remarkably,  given the freedom that the cubic splines have,  our reconstruction of $H(z)$ is close to the  $\Lambda$CDM prediction. 
 In this case, the odds compared to  the shape obtained using only BAO, $r^{\rm early}_{\rm s}$ and $H_0$  are 1:6, as this is an intermediate solution between the standard $\Lambda$CDM shape with low $H_0$ and the wiggly reconstruction obtained above.

\begin{figure}[h]
\minipage{0.35\textwidth}
\begin{center}
\includegraphics[width=\textwidth]{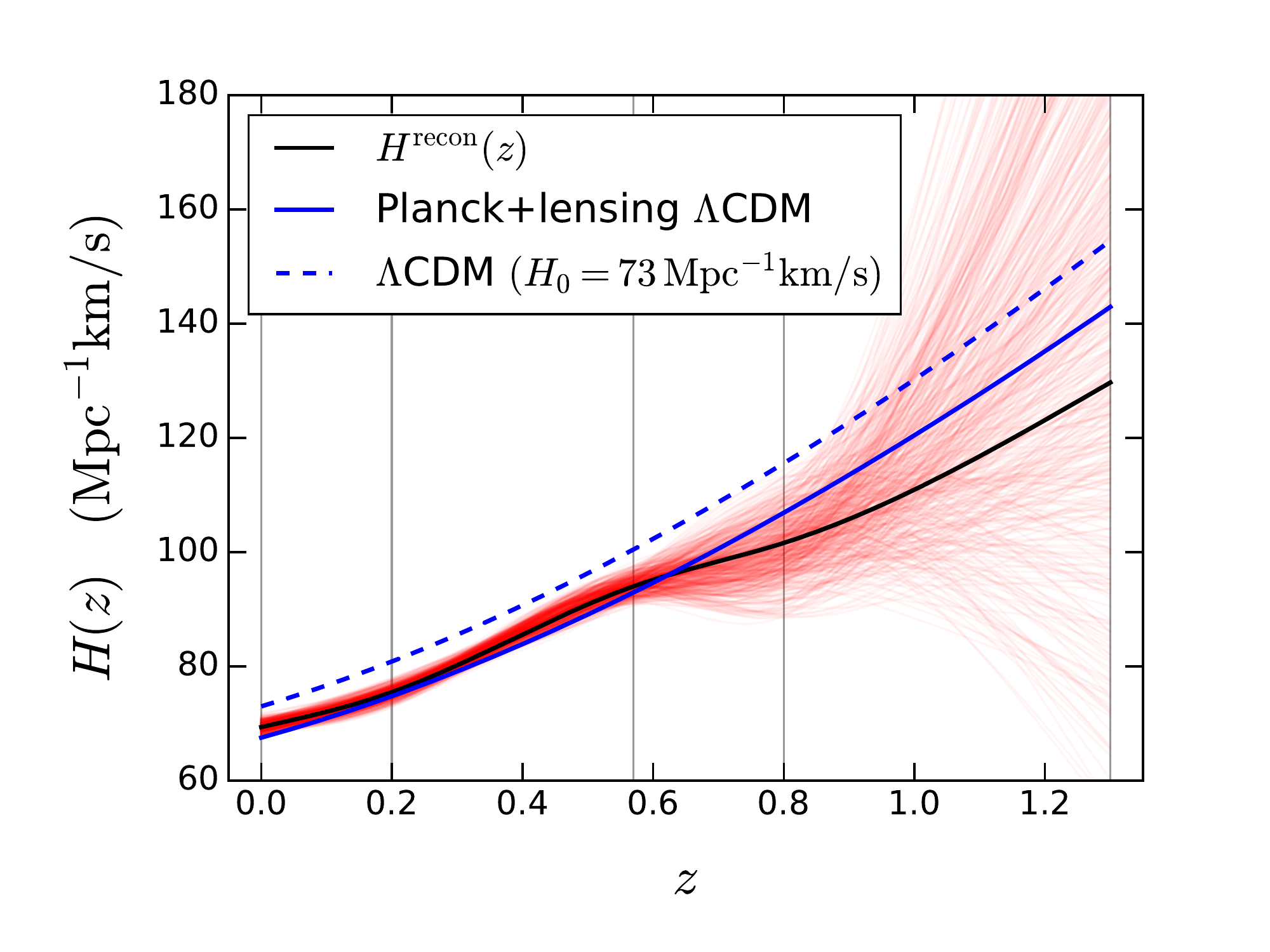}
\end{center}
\endminipage
\hspace{-0.5cm}
\minipage{0.35\textwidth}
\begin{center}
\includegraphics[width=\textwidth]{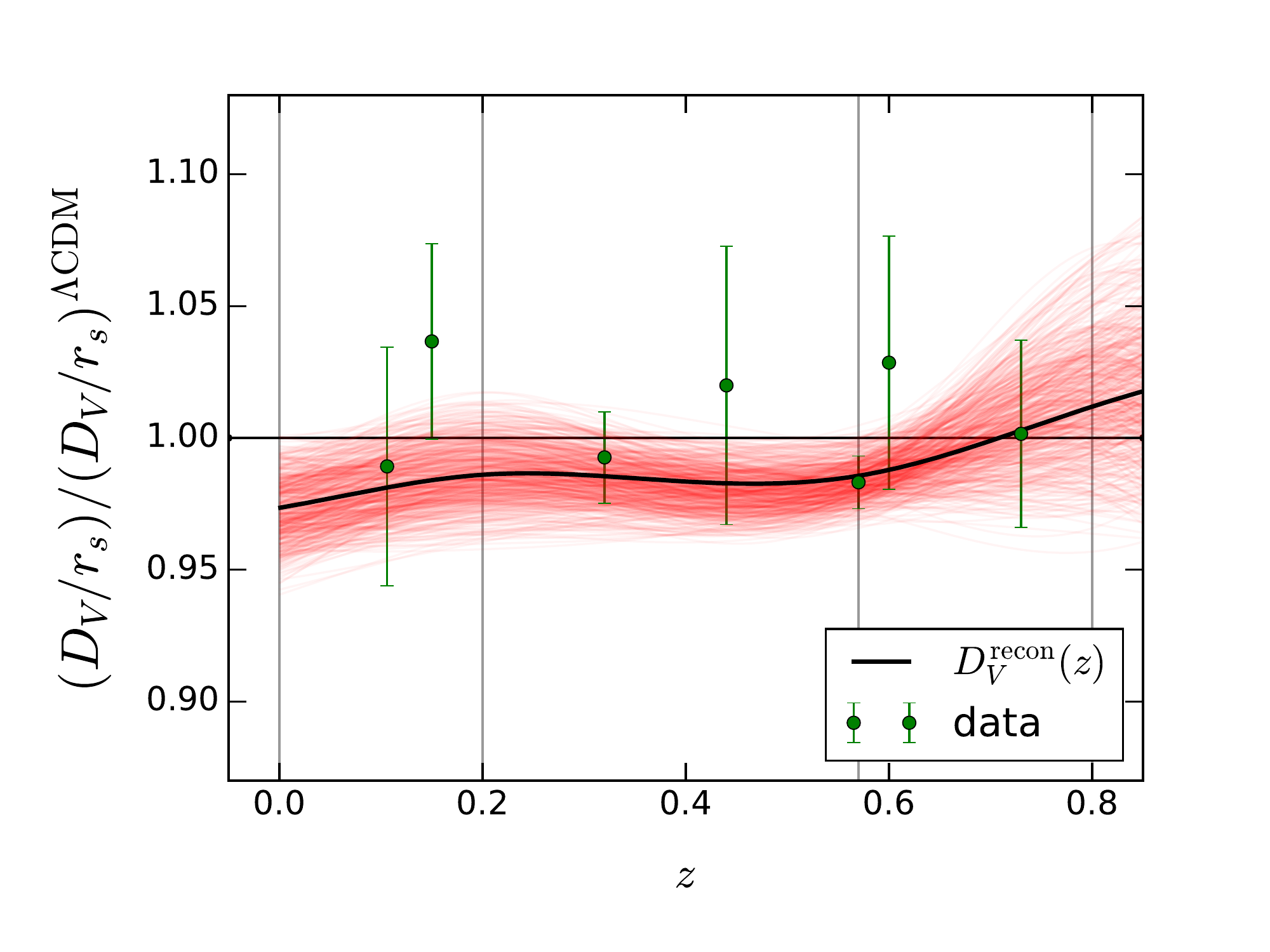}
\end{center}
\endminipage
\hspace{-0.5cm}
\minipage{0.35\textwidth}
\begin{center}
\includegraphics[width=\textwidth]{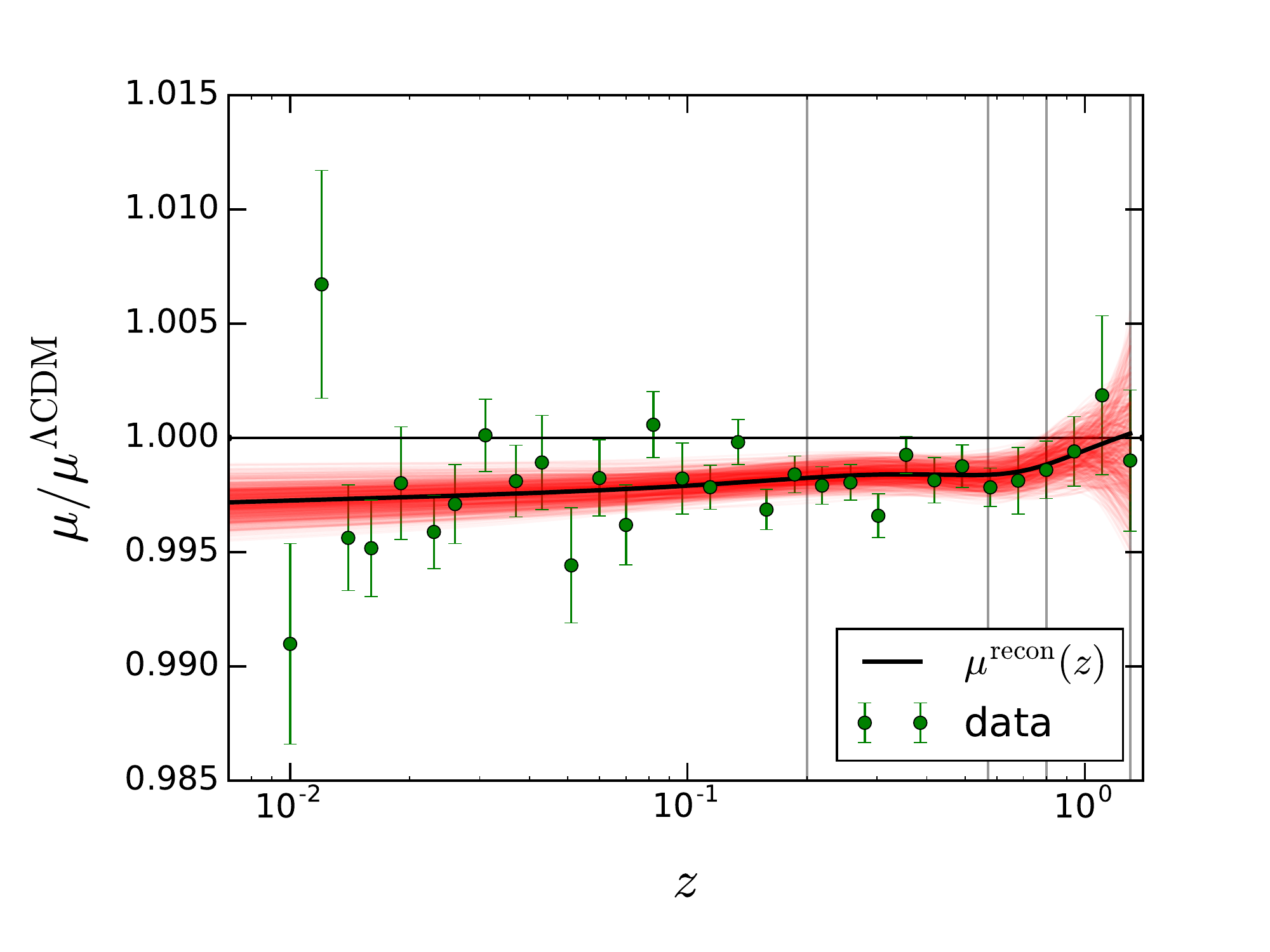}
\end{center}
\endminipage\hfill
\caption{\footnotesize
\textit{Left}: Results of the reconstruction of $H(z)$ using the direct measurement of $H_0$ of \citep{RiessH0_2016}, BAO and SNeIa data set with a CMB-derived $r_{\rm s}$ prior. 
\textit{Middle and right}: Observational data included in the reconstruction (green), as in the previous cases, and the prediction using the corresponding $H^{\rm recon}(z)$. 
}
\label{fig:H_recon_BAO+SN}
\end{figure}

To summarise, when using the local $H_0$ measurement and BAO data normalised to the CMB-derived $r_{\rm s}^{\rm early}$ under standard early Universe assumptions, the reconstruction indicates a sharp increase in the cosmic acceleration rate  ($H(z)\sim $constant) at $z<0.2$, where the BAO data have little statistical power. A dark energy equation of state parameter $w<-1$ (or dropping recently below $-1$) would fit the bill. However, when including SNeIa the shape of the  expansion history is constrained not to deviate significantly from that of $\Lambda$CDM (at $z<0.6$ where there are  many data points)  and thus only the  normalisation can  adjust, taking a  value  intermediate between the low and high redshift ``anchors", as $H_0r_{\rm s} \approx $ constant.
Thus a phantom dark energy is not favoured by the data. Below we will show that 
relaxing the flatness assumption does not change the results qualitatively.

\subsection{Reconstruction independent from the early-time physics}\label{sec:H0-rs}
To remove the dependence on the early Universe  assumptions introduced by the high redshift BAO anchor, we now treat  $r_{\rm s}$ as a free parameter in our analysis without including the prior of the early Universe of $r_{\rm s}^{\rm early}$. We consider the data set combination $H_0$, BAO(*) and SNeIa (hereafter the ``*" symbol indicate that no CMB-derived $r_{\rm s}^{\rm early}$ prior  is used).

\begin{figure}[h]
\begin{center}
\includegraphics[width=0.8\textwidth]{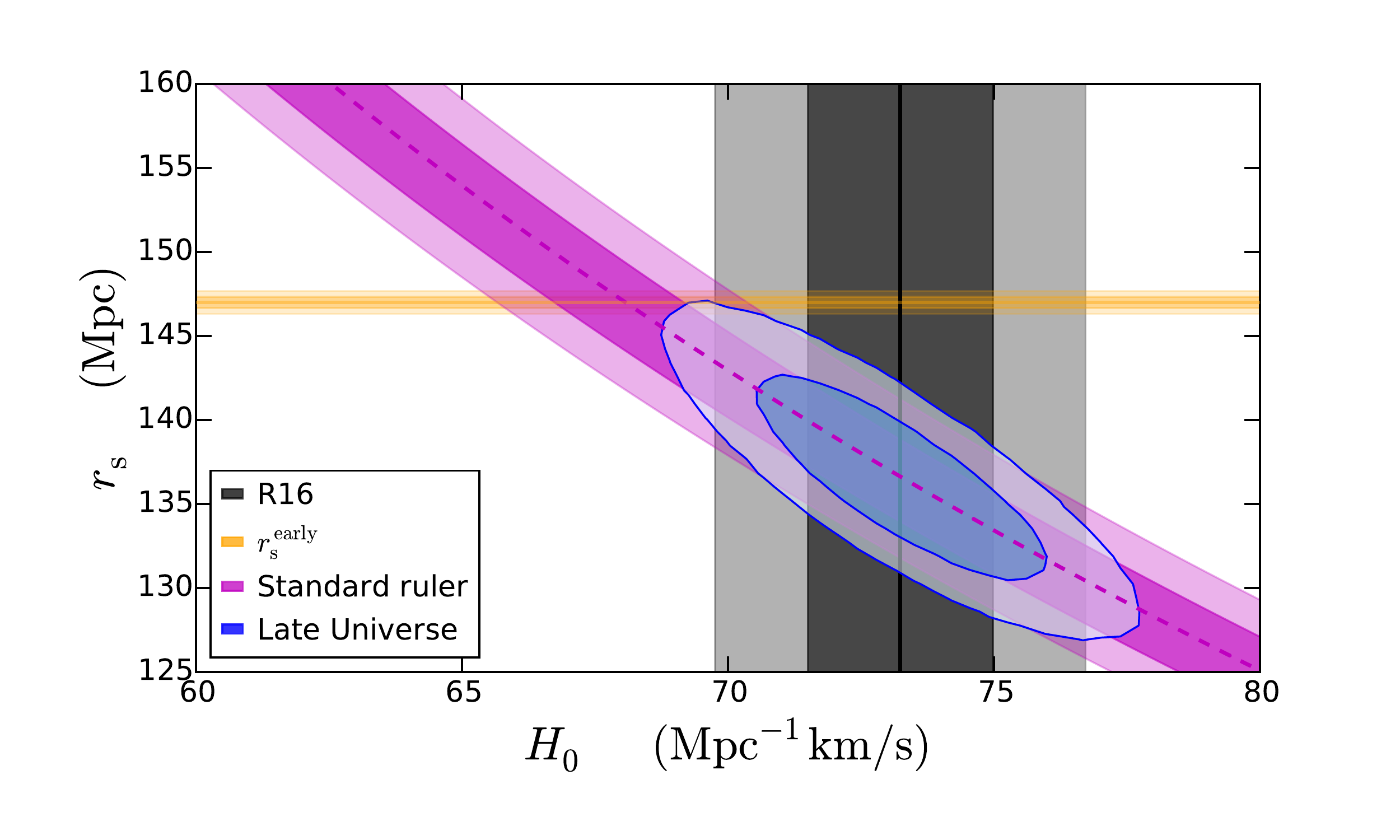}
\end{center}
\caption{\footnotesize
 Without an absolute distance scale, such as an $H_0$ determination, the low redshift standard ruler measurement  constrains the combination $r_{\rm s}h$,  and thus $r_{\rm s}H_0$ which appears as a (purple) band. This measurement does not include the prior on $r^{\rm early}_{\rm s}$. The local $H_0$ measurement (gray) can be used to break the degeneracy (yielding the blue confidence regions).}
\label{fig:pedagogical}
\end{figure}

This is illustrated in Fig.~\ref{fig:pedagogical}. The low redshift standard ruler measurement  constrains the combination $r_{\rm s}h$  which is reported here as a band in the $H_0$-$r_{\rm s}$ plane, constraining $H_0r_{\rm s} = $ constant. This constraint only relies on the BAO  yielding a standard ruler (of unknown length), on SNe being standard candles (of unknown luminosity), on spatial flatness and  on a smooth expansion history. The local $H_0$ measurement or the early-time $r_{\rm s}$ ``anchors" can be used to break the degeneracy. The $r_{\rm s}$ measurement relies on early-time physics assumptions (i.e. the value of $N_{\rm eff}$, $Y_{\rm P}^{\rm BBN}$, recombination physics,  epoch of matter-radiation equality etc.). The $H_0$  measurement relies on local calibrators of the cosmic distance ladder.

\begin{figure}[h]
\minipage{0.35\textwidth}
\includegraphics[width=\textwidth]{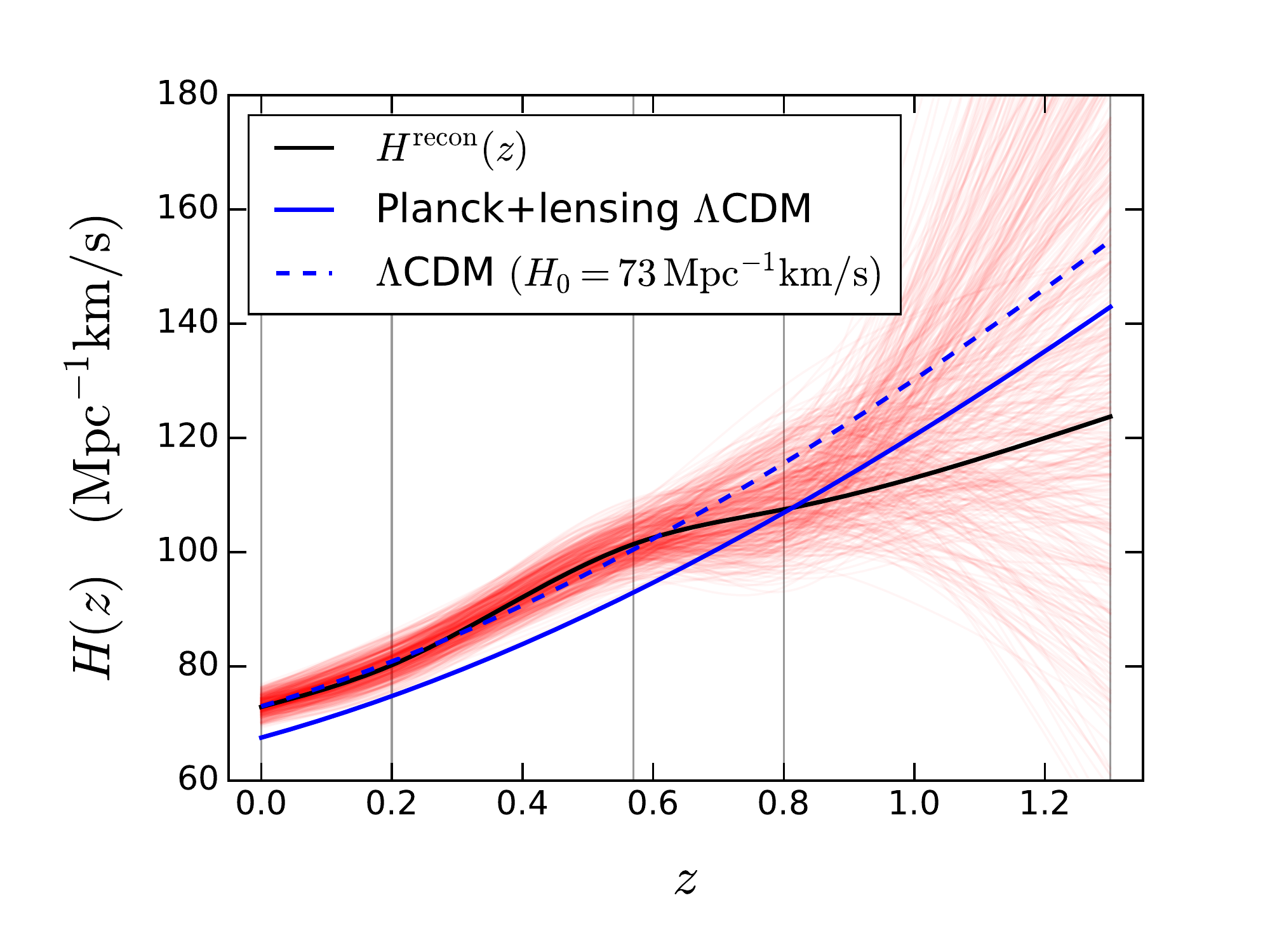}
\endminipage
\hspace{-0.5 cm}
\minipage{0.35\textwidth}
\includegraphics[width=\textwidth]{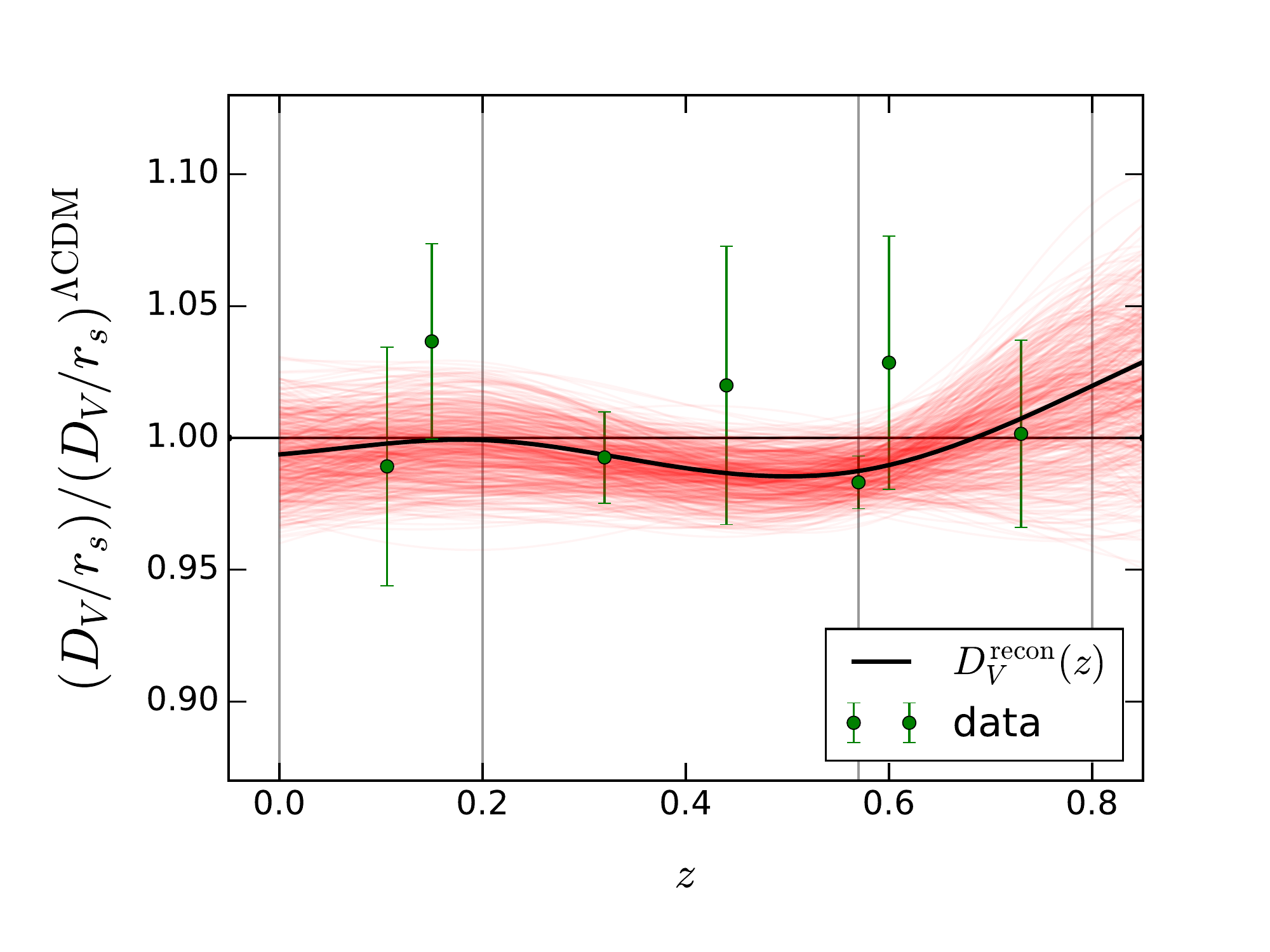}
\endminipage
\hspace{-0.5 cm}
\minipage{0.35\textwidth}
\includegraphics[width=\textwidth]{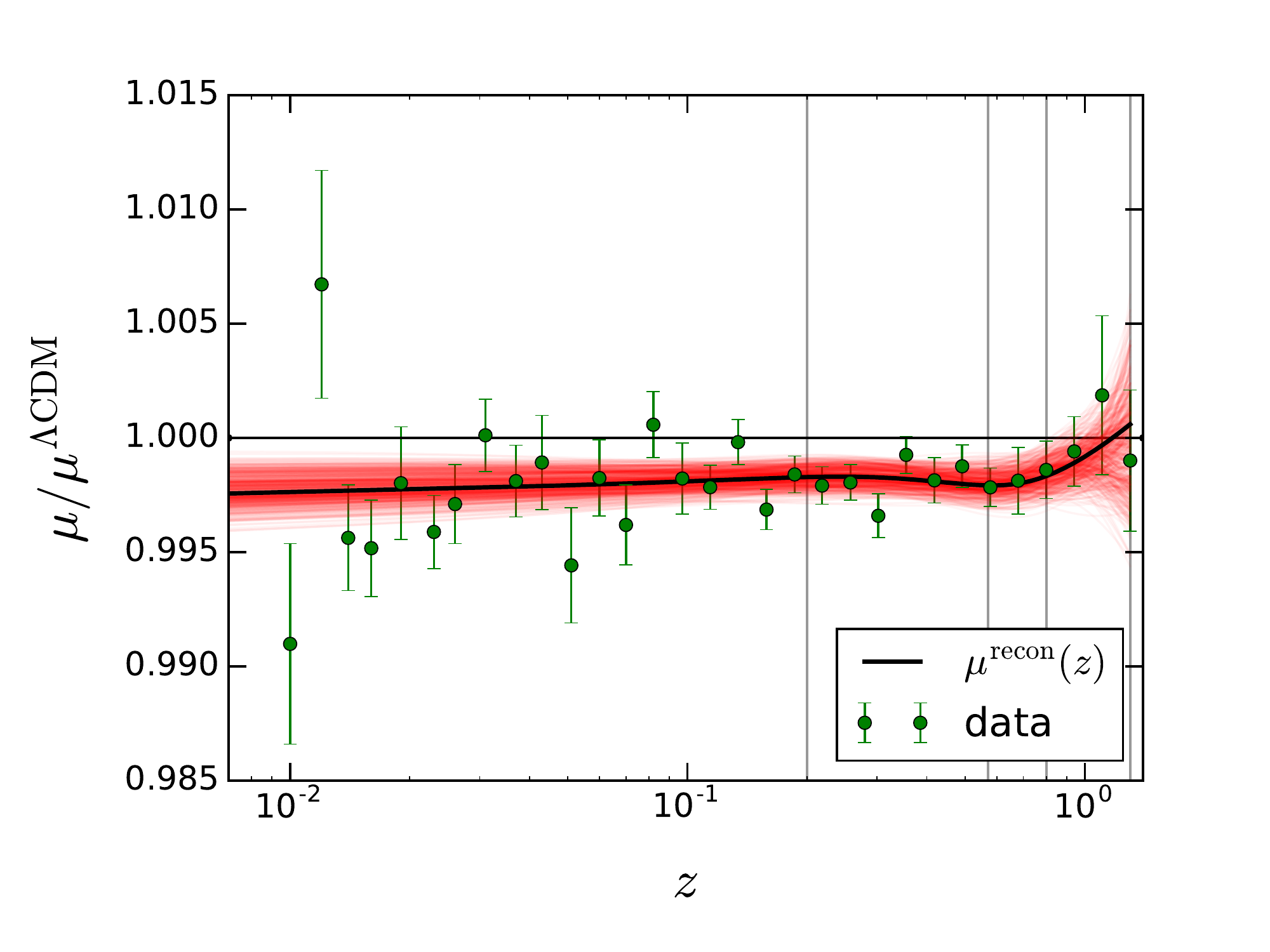}
\endminipage\hfill
\caption{\footnotesize
\textit{Left}: Reconstructed of $H(z)$ using the direct measurement of $H_0$ of \citep{RiessH0_2016}, BAO(*) and SNeIa data set and letting $r_{\rm s}$ vary as  a free parameter. 
\textit{Middle and right}: Observational data included in the reconstruction (green), as in previous cases, and the prediction using the corresponding $H^{\rm recon}(z)$.
}
\label{fig:H_recon_rs}
\end{figure}
 
The constraints on our parameters are reported in table \ref{tab:H_recon} and the reconstruction results are shown in figure \ref{fig:H_recon_rs}, using the same conventions as in previous plots. Once we free the  CMB-anchor of BAO (no $r_{\rm s}$ prior) the reconstruction  is similar to that with only SNeIa: $H^{\rm recon}(z)$ is very similar to the $\Lambda$CDM prediction (with $H_0 \sim 73$ $ {\rm Mpc^{-1}km/s}$).  $H^{\rm recon}(z=0)$ is in $\sim 2.9 \sigma$ tension with the value obtained by Planck 2015 \citep{Planckparameterspaper}, which assumes $\Lambda$CDM.  This procedure yields a model-independent estimate of  $r_{\rm s}=136.7 \pm 4.1$ Mpc; remarkably,  with an error  small enough to raise a tension of $2.6\sigma$ with Planck 2015-derived value assuming $\Lambda$CDM, and $2.5\sigma$ if we compare with $r_{\rm s}^{\rm early}$ of \citep{edepaper}.
This tension in $r_{\rm s}$ between the model independent measurement and the CMB-inferred value is entirely due to  the tension in $H_0$, via the relation $H_0r_{\rm s}=$ constant.
 As in previous cases, this data combination disfavours a recent sharp acceleration given by the shape of $H^{\rm recon}(z)$ using $H_0$ and BAO, with odds of 1:65.

\begin{figure}
\begin{center}
\includegraphics[width=0.75\textwidth]{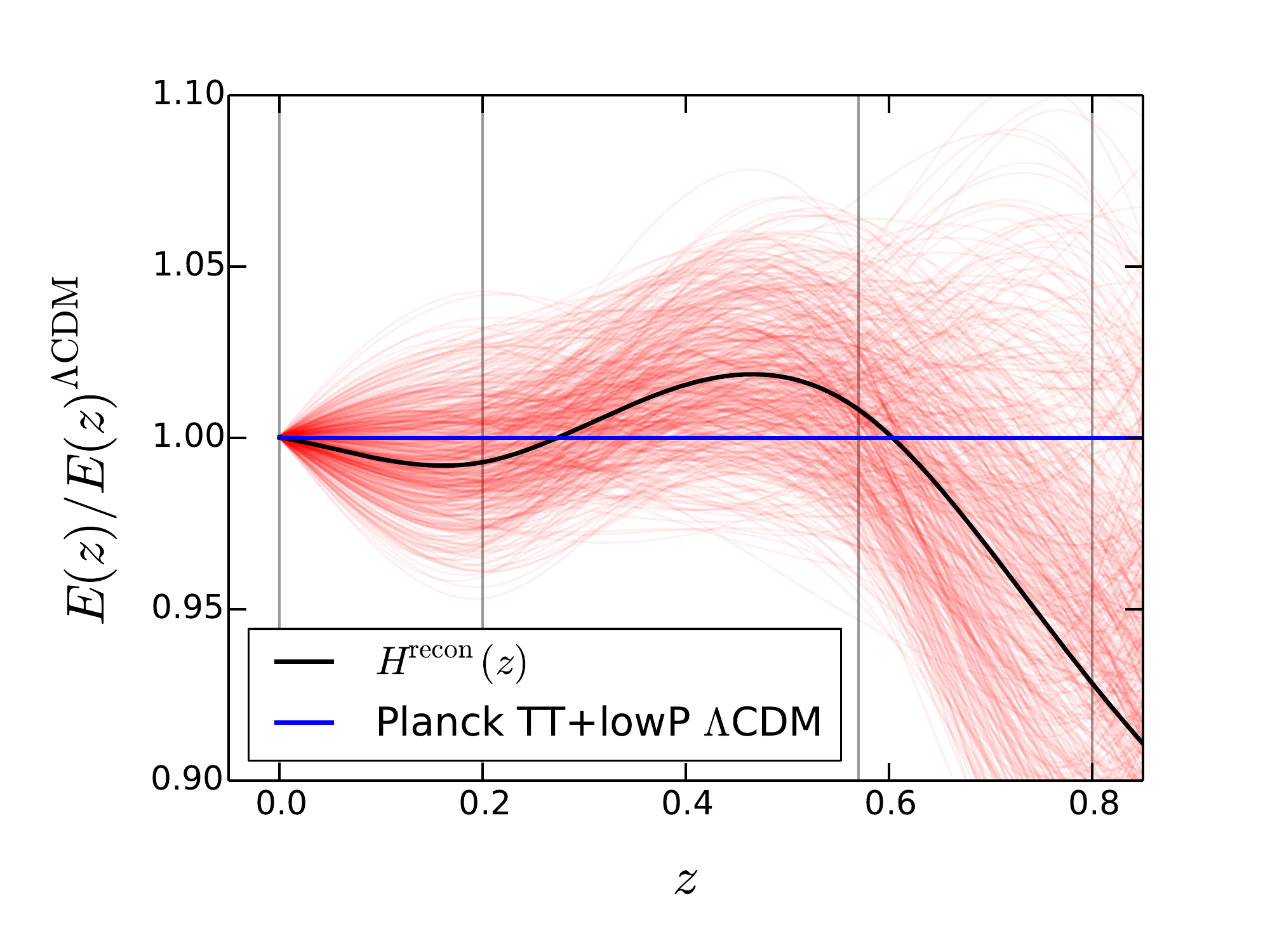}
\caption{\footnotesize Reconstructed shape of the expansion history  $E(z)=H(z)/H_0$, normalised to a standard $\Lambda$CDM expansion (with Planck-inferred density parameters), each with its corresponding value of $H_0$. At $z<0.6$ where the data have most statistical power, maximal deviations are at the 5\% level.}
\label{fig:H_recon_ratio}
\end{center}
\end{figure}

 It is interesting to note that the  reconstructed {\it shape} of $H^{\rm recon}(z)$ is constrained to be very close to the $\Lambda$CDM-predicted shape. This is illustrated in  figure \ref{fig:H_recon_ratio} for the combination SNeIa, BAO(*) and $H_0$: the maximal deviations at $z<0.6$, where the data have most of their statistical power, are at the 5\% level.

\begin{figure}

\minipage{0.85\textwidth}
\begin{center}
\includegraphics[width=0.75\textwidth]{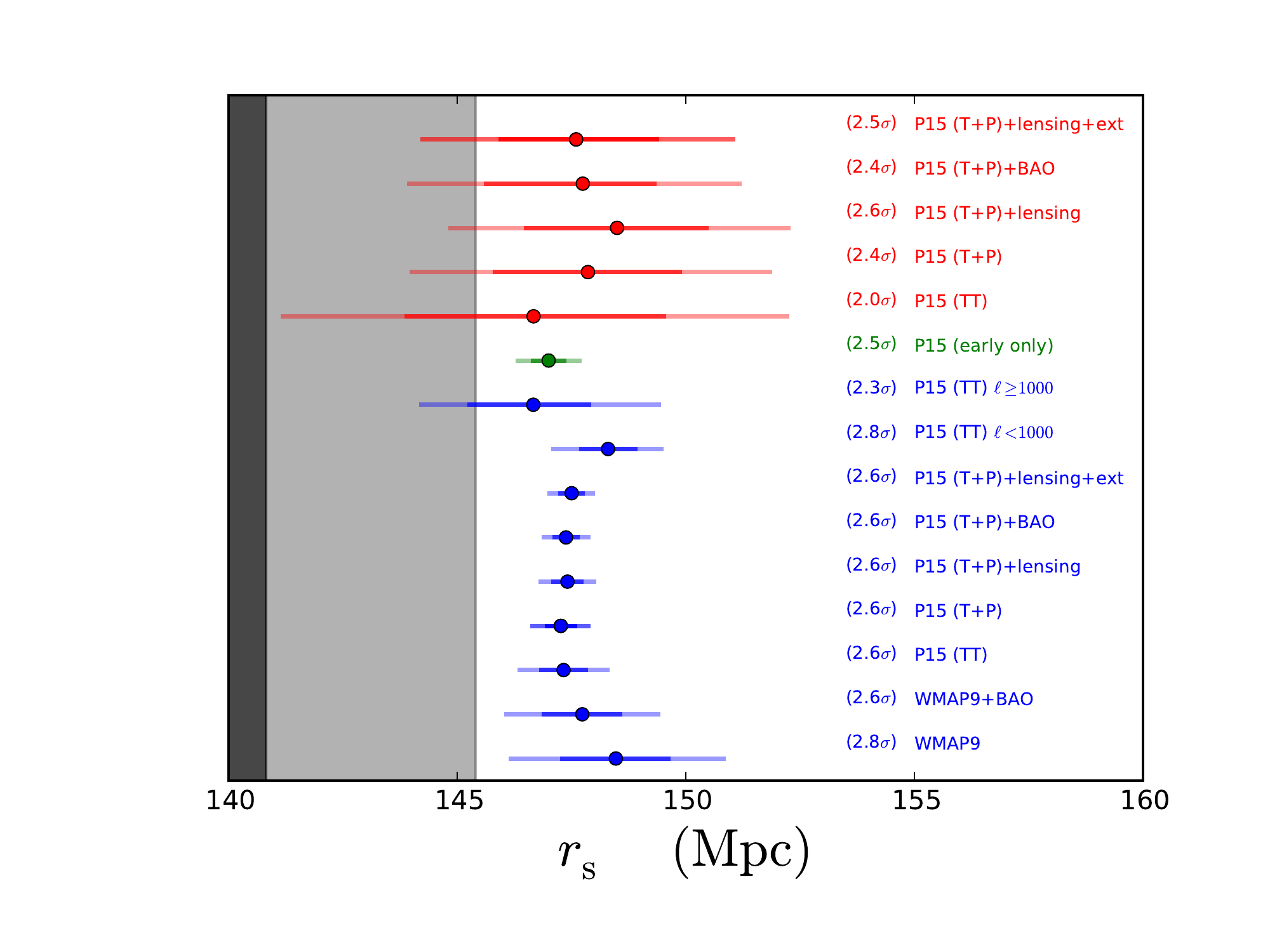}
\end{center}
\endminipage
\minipage{0.15\textwidth}
\hspace{-2cm}
\includegraphics[width=1.05\textwidth]{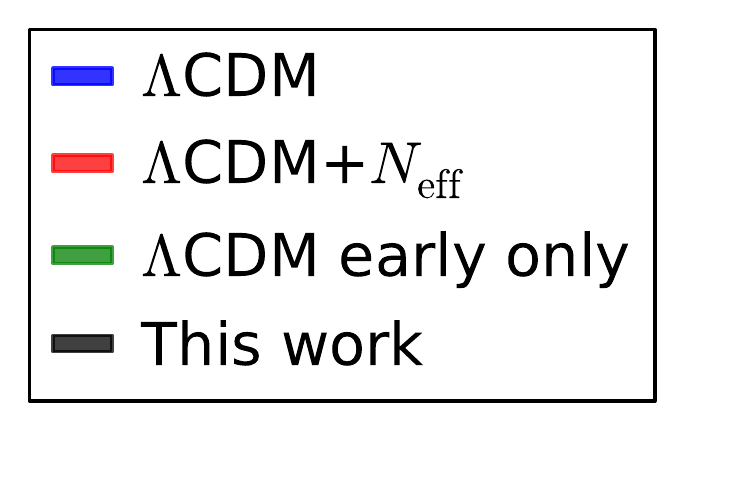}
\endminipage\hfill
\caption{\footnotesize
Marginalised 68\% and 95\% constraints on  $r_{\rm s}$ for different analysis of CMB data, obtained from Planck Collaboration 2015 public chains \citep{Planckparameterspaper}, WMAP9 \citep{HinshawWMAP_13} (analysed with the same assumptions than Planck) and the results of the work of  \citep{Addison_2016} (TT, $\ell>1000$ and $\ell<1000$) and  \citep{edepaper} (early only). We show  the constraints obtained in a $\Lambda$CDM context in blue,  $\Lambda$CDM+$N_{\rm eff}$ in red, the constraints obtained by analysing the CMB without any assumption of late universe physics (taken from \citep{edepaper}) in green. The result from this work is shown as a gray band. 
}
\label{fig:rsval}
\end{figure}

For completeness, in figure \ref{fig:rsval}, we show different CMB constraints on $r_{\rm s}$ compared with our measurement (black bands). Only when $N_{\rm eff}$ is free and only Planck temperature and lowP data are used, the constraint is modestly consistent ($\sim 2\sigma$). In the other cases,  the tension is significant, except for  the analysis of Planck 2015 temperature power spectrum limited to $\ell \geq 1000$ \citep{Addison_2016}. However, as discussed by these authors,  most of the parameters  obtained  using only this $\ell$ range are in tension with Planck 2015 $\ell < 1000$ and WMAP9 constraints. \footnote{Note that the tensions in each case (in parenthesis in figure \ref{fig:rsval}) are very similar in most of the cases although by eye it may not be obviously apparent. This is because the error on the low redshift determination  is much larger and dominates the comparison.}

\begin{figure}
\minipage{0.5\textwidth}
\begin{center}
\includegraphics[width = 1.17\textwidth]{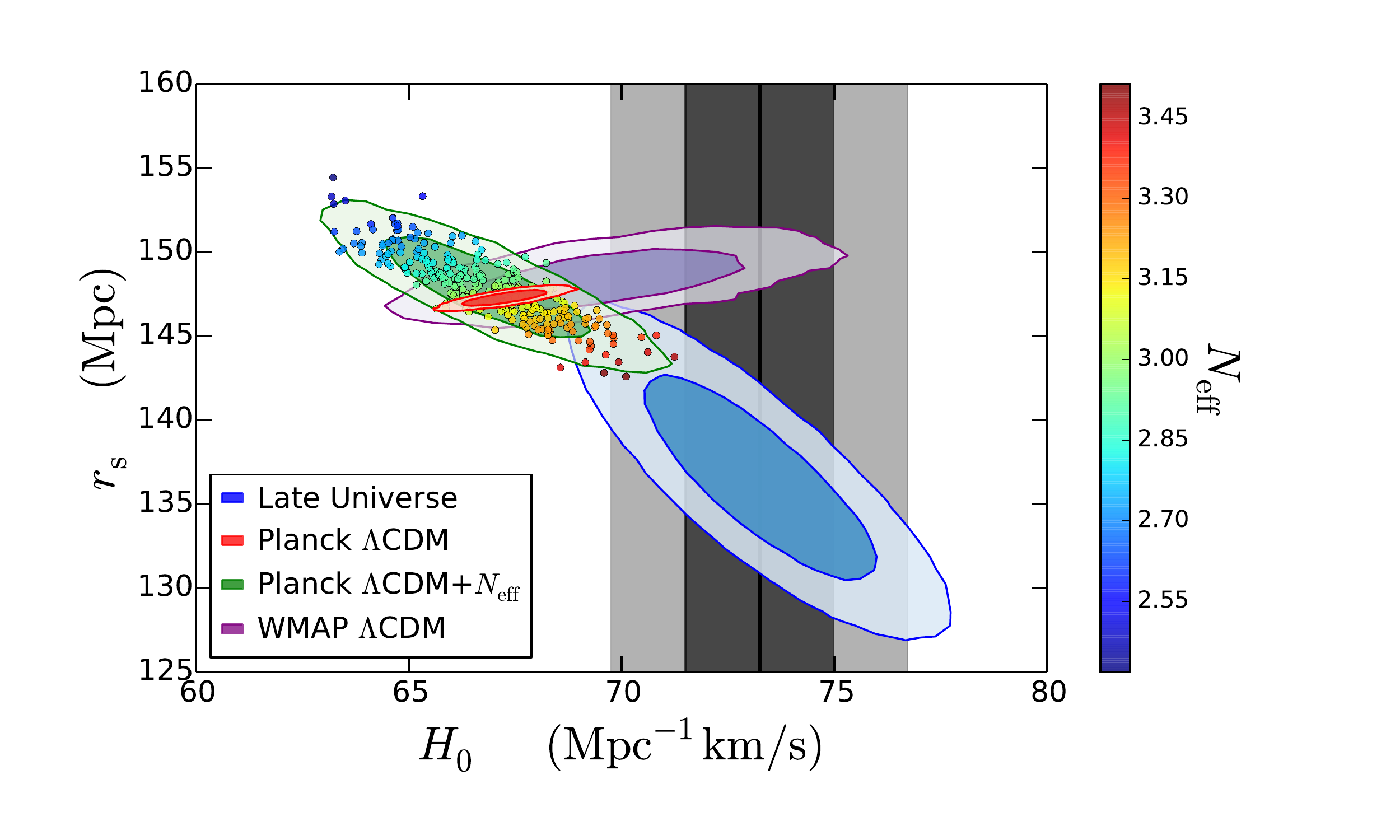}
\end{center}
\endminipage
\minipage{0.5\textwidth}
\begin{center}
\includegraphics[width = 1.17\textwidth]{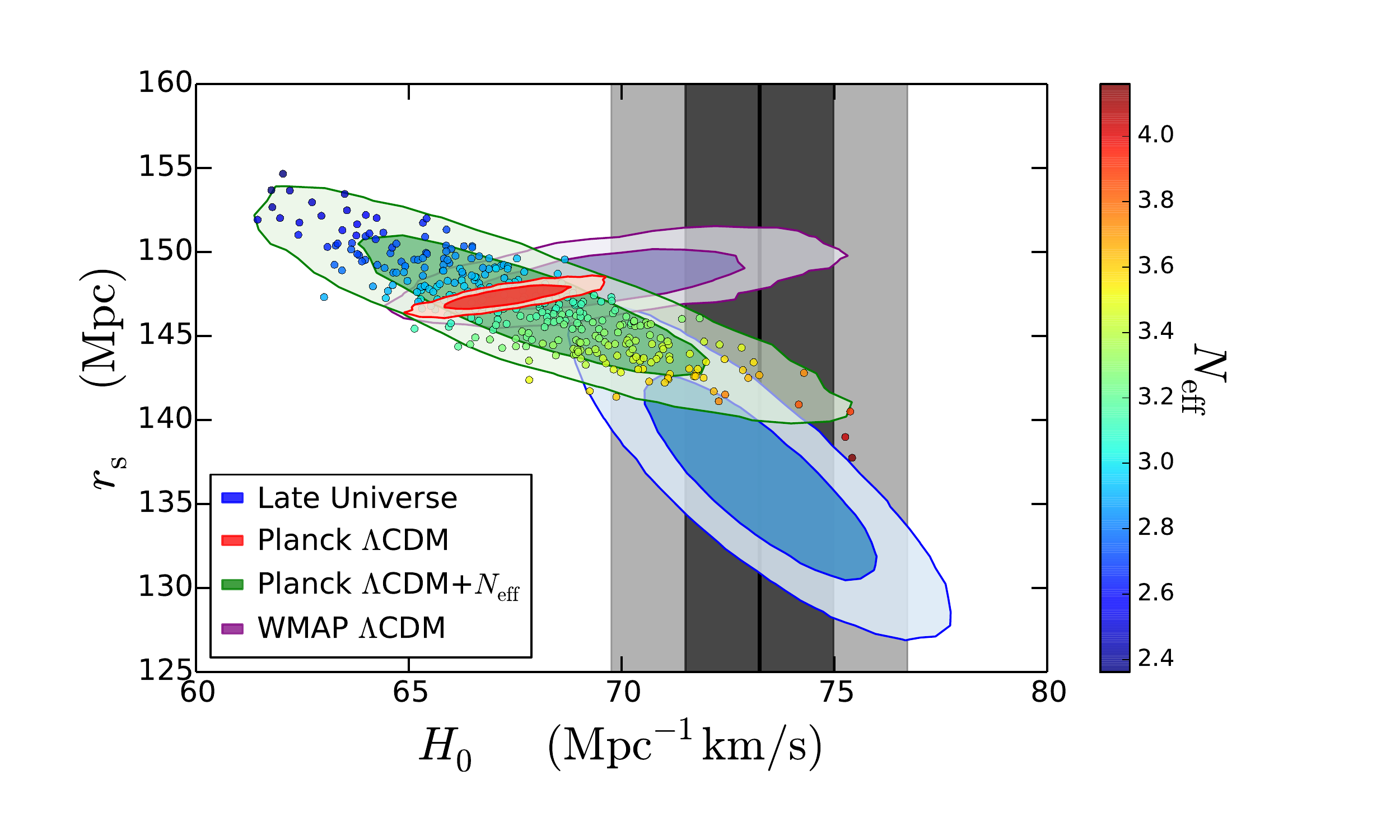}
\end{center}
\endminipage\hfill
\caption{\footnotesize
Marginalized constraints in the $H_0$-$r_{\rm s}$ plane  (68\% and 95\% regions)  for the cases discussed in the text. Planck data includes high $\ell$ polarization in the left panel and does not in the right one. Color-coded are  the corresponding values of $N_{\rm eff}$ in the case of $\Lambda$CDM+$N_{\rm eff}$.} 
\label{fig:H0-rs}
\end{figure}

It is illustrative to show the joint constraints for $H_0$ and $r_{\rm s}$, which we do in figure \ref{fig:H0-rs}. The vertical band is the local $H_0$ measurement and the blue contours are the constraints obtained in this work. As said before, $H_0$ and $r_{\rm s}$ are related by $H_0r_{\rm s}\approx$ constant and they are perfectly anticorrelated in our measurement. Here the perfect degeneracy is lifted by the measurement of $H_0$. In the same way, if the prior $r_{\rm s}^{\rm early}$ is included instead $H_0$ from \citep{RiessH0_2016}, the constraint on $H_0$ will be approximately\footnote{This relation is not exact because the prior $r_{\rm s}^{\rm early}$ is applied at various redshift and in combination with the BAO measurements.} $H_0^{\rm m}r_{\rm s}^{\rm m}/r_{\rm s}^{\rm early} \sim 68$ ${\rm Mpc^{-1}km/s}$, with the superscript {\rm m} meaning ``measured", recovering a value of $H_0$ close to the Planck-inferred value. 

We also  show the results of Planck using temperature and polarization power spectra (left) and only temperature power spectrum and lowP (right) for a   $\Lambda$CDM model (red) and a model which varies $N_{\rm eff}$ (green);  and of WMAP9  for a $\Lambda$CDM model (analysed in the same way as Planck, purple). It is possible to appreciate how, as CMB experiments derive  constraints assuming a model, the correlation is different and it depends strongly on the adopted  model (i.e $\Lambda$CDM vs. $\Lambda$CDM+$N_{\rm eff}$).

In table \ref{tab:tension2d}, we report the tension in the $H_0$-$r_{\rm s}$ plane between our measurement and CMB experiments for different models (computed as explained in section \ref{sec:Methods}), expressed as $\log \mathcal{T}$, the odds of full consistency and the tensions in terms of number of $\sigma$ (computed assuming gaussianity). Only for a model with extra dark radiation and discarding Planck's high $\ell$ polarisation data the two constraints  are in  acceptable agreement (i.e., the tension is not considered strong). 
From Fig.~\ref{fig:H0-rs} it is possible to appreciate that  within  $\Lambda$CDM,  excluding polarisation data makes the  CMB--derived $H_0$ value more consistent with the local $H_0$, but mainly because of an increase of the error bars;
however  including   or excluding polarisation data does not alter significantly the $r_{\rm s}$ determination.   To make the local $H_0$ determination, the low-redshift  estimate of the combination  $r_{\rm s}h$ and the CMB $r_{\rm s}$ determination fully consistent with each other, $r_{\rm s}$ should be significantly lowered. Among the $\Lambda$CDM extensions we explored,   the only one that   achieves this is  allowing   $N_{\rm eff}\sim 3.4$.

\begin{table}
\small
\begin{center}
\begin{tabular}{|c|c|c|c|c|}
\hline
	&	$\log \mathcal{T}$	& Odds & Jeffrey's modified scale	& Gaussian tension\\
\hline
Planck $	\Lambda$CDM (T+P) & 4.75 & 0.0086 (1:116) & strong & 3.7\\
Planck $	\Lambda$CDM (TT) & 4.31 & 0.013 (1:76) &strong &3.5\\
Planck $	\Lambda$CDM+$N_{\rm eff}$ (T+P) & 3.55 & 0.029 (1:34)&strong &3.1\\
Planck $	\Lambda$CDM+$N_{\rm eff}$ (TT) & 2.02 & 0.13 (1:8) & positive & 2.1\\
WMAP $\Lambda$CDM & 3.59 & 0.027  (1:37)& strong & 3.1 \\
\hline
\end{tabular}
\end{center}
\caption{Two dimensional tension between the low redshift joint constraints on $H_0-r_{\rm s}$ and a set of CMB-derived constraints. See section \ref{sec:Methods} for details.}
\label{tab:tension2d}
\end{table}

Finally, we also explore the case where the curvature of the Universe is not fixed to  $\Omega_{\rm k}=0$.In this case, $\Omega_{\rm k}$ remains largely unconstrained, and still broadly consistent with zero.
There is no significant shift in the rest of parameters, but the error bars are larger. The constraints are summarized in the bottom rows of table \ref{tab:H_recon}. Given the freedom of our expansion history reconstruction, to obtain useful constraints on $\Omega_{\rm k}$ more data would be needed (see  for example, \cite{Heavens:2014rja,StandardQuantities} where   curvature constraints  are reported). In the near future, anisotropic measurements of the BAO feature from on-going and forthcoming surveys could also be used.

\section{Discussion and Conclusions}\label{sec:Conclusions}

The standard $\Lambda$CDM model  with only a handful of  parameters,  provides an excellent description of a host of cosmological observations with remarkably few exceptions.   The most notable and persistent one is the local determination of the Hubble constant $H_0$, which, with the recent improvement by \citep{RiessH0_2016}, presents a  $\sim 3 \sigma$ tension with respect to the value inferred by the Planck Collaboration (assuming $\Lambda$CDM).   The CMB is mostly sensitive to early-Universe physics, and the CMB-inferred $H_0$ measurement thus depends on assumptions about both early time and late-time physics. A related quantity that the CMB can measure in a way that does not depend on late-time physics is the sound horizon at radiation drag, $r_{\rm s}$. This measurement however is  still model-dependent in that it relies on  standard assumptions about early-time physics.  On the other hand the local measurement of $H_0$ is model-independent as it does not depend on cosmological assumptions. As this work was nearing completion, new quasar time-delay  cosmography data became available \cite{H0_holicow}. Within the $\Lambda$CDM model  these provide an $H_0$ constraint  centered around 72 ${\rm Mpc}^{-1}$km/s,  with a ~4\% error and thus shows  reduced tension.

The two parameters $r_{\rm s}$ and $H_0$ are strictly related when we consider   also BAO observations. Expansion history probes such as BAO  and SNIa can provide a model-independent estimate of  the  low-redshift standard ruler, constraining directly  the combination $r_{\rm s}h$ (with $H_0=h \times 100$ ${\rm Mpc}^{-1}$km/s).  Thus $r_{\rm s}$ and $H_0$  provide  absolute scales for distance measurements  (anchors) at opposite ends of the observable Universe.  In the absence of systematic errors in the measurements,   if the standard cosmological model is the correct model, indirect (model-dependent) and direct (model-independent) constraints on these parameters should agree. The tension could thus provide evidence of physics beyond the standard model (or unaccounted systematic errors).

We have performed a complete cosmological study of the current tension between the inferred value of $H_0$ from the latest CMB data (as provided by the Planck satellite) \citep{Planckparameterspaper} and its direct measurement, with the recent update from \citep{RiessH0_2016}.  This reflects into a tension between cosmological model-dependent and model-independent constraints on $r_{\rm s}$.

We first  have explored  models for deviations from the standard $\Lambda$CDM in the early-Universe physics. 
When including CMB data alone (or  in combination with geometric measurements that do not rely on the $H_0$ anchor such as BAO) we find no evidence for deviations from the standard $\Lambda$CDM model and in particular no evidence for extra  effective relativistic species beyond three active neutrinos. This conclusion is unchanged if we allow additional freedom in the behaviour of the perturbations, both in  all relativistic species or only in the additional ones.

Therefore we put limits on the possible presence of a Universe  component whose mean energy scales like radiation with the Universe expansion but which perturbations could behave like radiation,  a perfect fluid,  a scalar field or anything else in between. 
On the other hand the value for the Hubble constant inferred by these analyses and other promising modifications of early-time physics, is always significantly lower than the local measurement of~\cite{RiessH0_2016}.  Should the  low-level systematics present in the  high $\ell$  ``preliminary" Planck polarisation data be found  to be non-negligible, the TEEE data should not be included in the analysis. In this case,  including only the ``recommended" baseline of low $\ell$ temperature and polarisation data and only temperature for  high $\ell$,  the tight limits relax and the tension disappears for a cosmological model with extra dark radiation corresponding to  $\Delta N_{\rm eff} \sim 0.4$. However the tension appears (but at an acceptable level) again when BAO data is included. The constraints on the effective parameters which describe the behaviour of the extra radiation in terms of perturbations are too weak to discriminate among the different candidates.

Another possible way to reconcile the CMB-derived $H_0$ value and the local measurement is to allow deviations from the standard  late-time expansion history of the Universe. Rather than invoking specific models  we have reconstructed the expansion history in a model-independent, minimally parametric way. Our method to reconstruct $H(z)$ does not rely on any model and only require minimal assumptions. These are: SNeIa form a homogeneous group and can be used as standard candles, $r_{\rm s}$ is a standard ruler for BAO corresponding to the sound horizon at radiation drag,  the  expansion history is smooth and continuous and  the Universe is spatially  flat.
When only using BAO, and the $H_0$ measurement with an early Universe $r_{\rm s}$ prior, the reconstructed $H(z)$ shows a sharp increase in acceleration at low redshift, such as that provided by a phantom equation of state parameter for dark energy. However when SNeIa are included, the shape of $H(z)$ cannot deviate significantly from that of a $\Lambda$CDM, disfavouring therefore the phantom dark energy solution. When the CMB $r_{\rm s}$  prior is removed, this procedure yields a model-independent determination of $r_{\rm s}$ (and  the expansion history)  without any assumption on the early Universe. The $r_{\rm s}$ value so obtained is significantly lower than  that obtained from the CMB assuming standard early-time physics (2.6$\sigma$ tension). 
When we relax the assumption about the flatness of the Universe,  the curvature remains largely unconstrained and the error on the other parameters grow slightly.  We do not find significant shifts in the rest of the parameters.

Of course this hinges on identifying the BAO standard ruler with the sound horizon at radiation drag. Several processes have been proposed that could displace the BAO feature, the most important being non-linearities, bias  e.g., \cite{Angulo, Rasera} and  non-zero baryon-dark matter  relative velocity \cite{TH2010, D2010, Slepian}.  These effects however have been found to be  below  current errors \cite{PB09, Blazek, Slepian2} and below the 1\% level. It is therefore hard to imagine how these effects could introduce the $~ 5-7 $\%  shift required to solve the tension.

In summary, because the shape of the expansion history is tightly constrained by current data, in a model--independent way,  the $H_0$ tension can be restated as a mis-match in the normalisation of the cosmic distance ladder between the two anchors: $H_0$ at low redshift and $r_{\rm s}$ at high redshift. In the absence of systematic errors, especially in the high $\ell$ CMB polarisation data and/or in the local $H_0$ measurement,  the mismatch suggest   reconsidering the  standard assumptions about early-time physics.
Should the  ``preliminary" high $\ell$ CMB polarisation data be found to be affected by significant systematics and excluded from the analysis, the mismatch could be resolved by allowing an extra component behaving like dark radiation at the background level with a $\Delta N_{\rm eff} \sim 0.4$.  Other new physics in the early Universe that reduce the CMB-inferred sound horizon at radiation drag by $\sim 10$ Mpc  (6\%) would have the same effect. 

\acknowledgments
We thank  Graeme Addison, Alan Heavens and Antonio J. Cuesta for valuable discussion during the development of this study, which helped to improve this work.
JLB is supported by the Spanish MINECO under grant BES-2015-071307, co-funded by the ESF. Funding for this work was partially provided by the Spanish MINECO under projects  AYA2014-58747-P and MDM-2014- 0369 of ICCUB  (Unidad de Excelencia Maria de Maeztu). JLB acknowledges hospitality of Radcliffe Institute for Advanced Study, Harvard University.

Based on observations obtained with Planck (http://www.esa.int/Planck), an ESA science mission with instruments and contributions directly funded by ESA Member States, NASA, and Canada.

Funding for SDSS-III has been provided by the Alfred P. Sloan Foundation, the Participating Institutions, the National Science Foundation, and the U.S. Department of Energy Office of Science. The SDSS-III web site is http://www.sdss3.org/.

SDSS-III is managed by the Astrophysical Research Consortium for the Participating Institutions of the SDSS-III Collaboration including the University of Arizona, the Brazilian Participation Group, Brookhaven National Laboratory, Carnegie Mellon University, University of Florida, the French Participation Group, the German Participation Group, Harvard University, the Instituto de Astrofisica de Canarias, the Michigan State/Notre Dame/JINA Participation Group, Johns Hopkins University, Lawrence Berkeley National Laboratory, Max Planck Institute for Astrophysics, Max Planck Institute for Extraterrestrial Physics, New Mexico State University, New York University, Ohio State University, Pennsylvania State University, University of Portsmouth, Princeton University, the Spanish Participation Group, University of Tokyo, University of Utah, Vanderbilt University, University of Virginia, University of Washington, and Yale University.

\providecommand{\href}[2]{#2}\begingroup\raggedright\endgroup
\bibliographystyle{utcaps}

\end{document}